\documentclass[conference]{IEEEtran}
\IEEEoverridecommandlockouts
\usepackage{amsmath,amsfonts}
\usepackage{mathrsfs}
\usepackage{bm}
\usepackage{nicefrac}
\usepackage{booktabs}
\usepackage{array}
\usepackage{multirow}
\usepackage{threeparttable}
\usepackage{makecell}
\usepackage[procnumbered,ruled,vlined,linesnumbered]{algorithm2e}
\usepackage{siunitx}
\usepackage{stfloats}
\usepackage{subfigure}
\usepackage{array}
\usepackage{balance}
\usepackage{tabularx}
\usepackage{lipsum}
\usepackage{xcolor}
\usepackage{tikz,tikz-cd}
\usepackage{pgfplots,capt-of}
\usepackage{graphicx}
\usepackage{textcomp}

\pgfplotsset{compat=1.18}

\newcommand{\inlcomm}[1]{}

\newcommand{\spar}[1]{\left( #1 \right)}
\newcommand{\normspar}[1]{( #1 )}
\newcommand{\midpar}[1]{\left[ #1 \right]}

\newcommand{\myord}[1]{{#1}^{\rm{th}}}
\newcommand{\setof}[1]{\left\{#1 \right\}}
\newcommand{\abs}[1]{\left|#1 \right|}
\newcommand{\mean}[1]{\mathbb{E}\midpar{#1}}
\newcommand{\ceil}[1]{\left\lceil#1\right\rceil}

\newcommand{\Abs}[1]{\left\Vert#1\right\Vert}

\newcommand{\diag}[1]{\mathrm{diag}\spar{#1}}
\newcommand{\inv}[1]{{#1}^{-1}}
\newcommand{\parinv}[1]{\inv{\spar{#1}}}
\newcommand{\bigparinv}[1]{\bigspar{#1}^{-1}}

\newcommand{\absorb}[1]{H_{#1}}
\newcommand{\cfcc}[1]{C(#1)}
\newcommand{\resist}[2]{R{({#1},{#2})}}

\newcommand{\edge}[2]{\spar{#1,#2}}
\newcommand{\algname}[1]{\textsc{#1}}
\newcommand{\bernfunc}[1]{f\spar{#1}}

\newcommand{\bigo}[1]{O\bigspar{#1}}
\newcommand{\tildeo}[1]{\tilde{O}\normspar{#1}}

\newcommand{\bigspar}[1]{\big( #1 \big)}
\newcommand{\biggspar}[1]{\bigg( #1 \bigg)}

\newcommand{\bigabs}[1]{\big| #1 \big|}

\newcommand{\bigtrace}[1]{\mathrm{Tr}\bigspar{#1}}

\newcommand{\normo}[1]{O\normspar{#1}}

\newcommand{\gr}{G}
\newcommand{\tilgr}{\tilde{\gr}}
\newcommand{\kem}{K}

\newcommand{\rea}{\mathbb{R}}
\newcommand{\rme}{\mathrm{e}}
\newcommand{\dmax}{d_{\max}}

\newcommand{\tillap}{\tilde{\lap}}

\newcommand{\gain}[2]{\Delta(#1,#2)}
\newcommand{\approxgain}[2]{\Delta'(#1,#2)}
\newcommand{\treemark}{N}
\newcommand{\forest}{\mathcal{F}}
\newcommand{\forestmark}{\mathcal{N}}
\newcommand{\tildeforestmark}{\tilde{\mathcal{N}}}
\newcommand{\treenum}[4]{{\treemark}_{#1,#2}^{{#3}\to{#4}}}
\newcommand{\forestnum}[4]{{\forestmark}_{#1,#2}^{{#3}\to{#4}}}
\newcommand{\forestfreq}[4]{{\tildeforestmark}_{#1,#2}^{{#3}\to{#4}}}
\newcommand{\pathway}[2]{\mathcal{P}_{{#1},{#2}}}
\newcommand{\voltage}[3]{\Phi_{#1,#3}(#2)}
\newcommand{\meanvolt}[3]{\overline{\Phi}_{#1,#3}(#2)}
\newcommand{\meanvolts}[3]{\overline{\varPhi}_{#1,#3}(#2)}
\newcommand{\rsf}[1][S]{\mathcal{F}_{#1}}
\newcommand{\parent}[1]{{\pi}_{#1}}
\newcommand{\dfslist}{\mathcal{L}_{\textnormal{DFS}}}
\newcommand{\bfslist}{\mathcal{L}_{\textnormal{BFS}}}
\newcommand{\complset}[1][S]{V\setminus{#1}}
\newcommand{\sublap}[1][S]{\lap_{-{#1}}}
\newcommand{\invsublap}[1][S]{\lap_{-{#1}}^{-1}}
\newcommand{\sqinvsublap}[1][S]{\lap_{-{#1}}^{-2}}
\newcommand{\schcompl}[2]{\mathcal{S}_{\scriptscriptstyle #2}(#1)}
\newcommand{\approxschcompl}[2]{\tilde{\mathcal{S}}_{\scriptscriptstyle #2}(#1)}
\newcommand{\matentry}[3]{#1_{{#2}{#3}}}
\newcommand{\invmatentry}[3]{#1_{{#2}{#3}}^{-1}}
\newcommand{\rootprob}[2]{\Pr\spar{\rootnode{#1}={#2}}}

\newcommand{\rootnum}[2]{{\tildeforestmark}\spar{\rootnode{#1}={#2}}}
\newcommand{\rootnode}[1]{\rho_{#1}}
\newcommand{\randgr}[2]{{#1}^{(#2)}}

\newcommand{\veca}{\boldsymbol{a}}
\newcommand{\vecb}{\boldsymbol{b}}
\newcommand{\vecc}{\boldsymbol{c}}
\newcommand{\vecd}{\boldsymbol{d}}
\newcommand{\vece}{\boldsymbol{e}}

\newcommand{\vecv}{\boldsymbol{v}}
\newcommand{\vecw}{\boldsymbol{w}}

\newcommand{\vecx}{\boldsymbol{x}}

\newcommand{\vecone}{\boldsymbol{1}}

\newcommand{\lap}{\boldsymbol{L}}
\newcommand{\mata}{\boldsymbol{A}}
\newcommand{\matb}{\boldsymbol{B}}
\newcommand{\matc}{\boldsymbol{C}}
\newcommand{\matd}{\boldsymbol{D}}
\newcommand{\matf}{\boldsymbol{F}}
\newcommand{\tildematf}{\tilde{\boldsymbol{F}}}

\newcommand{\mati}{\boldsymbol{I}}

\newcommand{\matm}{\boldsymbol{M}}
\newcommand{\matp}{\boldsymbol{P}}

\newcommand{\matq}{\boldsymbol{Q}}

\newcommand{\matw}{\boldsymbol{W}}

\newcommand{\maty}{\boldsymbol{Y}}

\newcommand{\lemref}[1]{Lemma~\ref{#1}}
\newcommand{\thmref}[1]{Theorem~\ref{#1}}

\newcommand{\algref}[1]{Algorithm~\ref{#1}}
\newcommand{\defref}[1]{Definition~\ref{#1}}
\newcommand{\secref}[1]{Section~\ref{#1}}
\newcommand{\tabref}[1]{Table~\ref{#1}}
\newcommand{\figref}[1]{Figure~\ref{#1}}

\newcommand{\lineref}[1]{Line~\ref{#1}}
\newcommand{\linerangeref}[2]{Lines~\ref{#1}-\ref{#2}}

\newcommand{\nicehalf}{\nicefrac{1}{2}}

\DeclareMathOperator*{\argmin}{arg\,min}
\DeclareMathOperator*{\argmax}{arg\,max}

\DontPrintSemicolon
\SetKw{KwAnd}{and}
\SetFuncSty{textsc}
\SetKwInOut{Input}{Input\ \ \ \ }

\SetKwInOut{Output}{Output}

\newtheorem{theorem}{Theorem}[section]

\newtheorem{lemma}[theorem]{Lemma}
\newtheorem{definition}[theorem]{Definition}

\renewcommand{\figref}[1]{Fig.~\ref{#1}}

\usepackage{comment}
\usepackage{amssymb}
\usepackage{url}
\usepackage[T1]{fontenc}
\def\BibTeX{{\rm B\kern-.05em{\sc i\kern-.025em b}\kern-.08em
    T\kern-.1667em\lower.7ex\hbox{E}\kern-.125emX}}

\begin{document}

\title{Fast Maximization of Current Flow Group Closeness Centrality}

\author{
    \IEEEauthorblockN{Haisong Xia}
    \IEEEauthorblockA{
        School of Computer Science\\
        Fudan University\\
        Shanghai,200433,China\\
        hsxia22@m.fudan.edu.cn
    }
    \and
    \IEEEauthorblockN{Zhongzhi Zhang}
    \thanks{Zhongzhi~Zhang is the corresponding author.}
    \IEEEauthorblockA{
        School of Computer Science\\
        Fudan University\\
        Shanghai,200433,China\\
        zhangzz@fudan.edu.cn
    }
}

\maketitle

\begin{abstract}
    Derived from effective resistances, the current flow closeness centrality (CFCC) for a group of nodes measures the importance of node groups in an undirected graph with $n$ nodes. Given the widespread applications of identifying crucial nodes, we investigate the problem of maximizing CFCC for a node group $S$ subject to the cardinality constraint $|S|=k\ll n$. Despite the proven NP-hardness of this problem, we propose two novel greedy algorithms for its solution. Our algorithms are based on spanning forest sampling and Schur complement, which exhibit nearly linear time complexities and achieve an approximation factor of $1-\frac{k}{k-1}\frac{1}{\mathrm{e}}-\epsilon$ for any $0<\epsilon<1$. Extensive experiments on real-world graphs illustrate that our algorithms outperform the state-of-the-art method in terms of efficiency and effectiveness, scaling to graphs with millions of nodes.
\end{abstract}

\begin{IEEEkeywords}
    centrality, resistance distance, combinatorial optimization, spectral graph theory.
\end{IEEEkeywords}

\section{Introduction}

As a fundamental problem in the field of complex networks, the identification of crucial nodes has garnered widespread research interest~\cite{LiZhLiZhWa24,BaZh22}. Due to its importance in network analysis, this problem has been applied to various areas, including graph mining~\cite{ShDeCh21}, leader-follower opinion dynamics~\cite{ZhSuLiZh23}, and influence maximization~\cite{FePaCaVa23,GuFeZhWa23}. The most intuitive approach to solve this problem is to design a centrality that evaluates the importance of each node, thus transforming the identification problem into a computational one. Previous studies~\cite{BeKl15,XiXuZhZh25,WhSm03,BoDeRi16} have proposed various centralities to analyze the roles of nodes in a graph, with a large proportion focusing on closeness centrality. The closeness of a node is defined as the reciprocal of the sum of shortest path distances between it and all other nodes in the graph. Despite its advantage over local centralities, closeness centrality only considers the shortest paths. This sensitivity to noise potentially leads to counterintuitive results~\cite{BeWeLuMe16}.

To address the issue of sensitivity, current flow closeness centrality (CFCC)~\cite{BrFl05} is presented based on resistance distance in the corresponding electrical networks. Resistance distance is a crucial metric for solving diverse fundamental problems in the field of graph theory. Specifically, in the data management community, resistance distance is utilized to design effective graph systems and applications~\cite{QiDhTaPeWa21,ZhChSaZhZh13,ShMaWuCh14,SrDa14}. Many researchers also develop efficient approximation of resistance distance to facilitate data management tasks~\cite{PeLoYoGo21,YaTa23}.
Since CFCC is defined via resistance distance, it has been applied to analyze the vulnerability of power grids~\cite{CeDeMi18,GuBaUrZu13} and optimize the performance of conductive films~\cite{KiNa21}. Additionally, resistance distance considers contributions from all paths between node pairs, which mirrors the propagation of electrical spike signals in brain networks~\cite{LeCh24}. This property enables CFCC to identify active neurons in the analysis of brain networks~\cite{LiDeTaPaTiRoMa24}. In other types of networks, the current flow-based closeness also outperforms classical counterparts in identifying crucial nodes~\cite{BeWeLuMe16}.

While most previous studies focus on identifying individual nodes, many practical problems essentially require finding a group of nodes that is the most important among all node groups with capacity \(k\ll n\), where \(n\) denotes the number of all nodes. For instance, how to place resources on \(k\) peers in P2P networks for easy access  by others~\cite{GkMiSa06}. Meanwhile, placing sensors of wireless networks involves selecting an optimal subset of nodes for sampling physical signals such as radiation or temperature~\cite{KrSiGu08,RaChVe14}. Finally, the challenge of point cloud sampling~\cite{DiChWaBa20,ChTiFeVeKo17} lies in selecting a representative subset of points to preserve the geometric features for reconstruction.

Recently, the concept of current flow closeness centrality (CFCC) has been extended from individual nodes to node groups~\cite{LiPeShYiZh19}. Additionally, the problem of selecting a node group of size \(k\), aiming to maximize its CFCC, has been proposed. The state-of-the-art algorithm has been developed to approximately address it, which utilizes a fast Laplacian solver. However, its running time, while nearly linear in the number of edges, remains prohibitively long for large-scale networks with more than 10 million edges.

\begin{comment}
However, the memory requirements of this Laplacian solver are relatively high, rendering the algorithm impractical for large-scale networks with millions of edges. Furthermore, the sequential implementation of this Laplacian solver limits its scalability on multi-core CPU architectures.
\end{comment}

\textbf{Contributions.}
In light of this limitation, we establish the connection between Current Flow Closeness Maximization (CFCM) and rooted spanning forests, motivating our proposal of a greedy Monte Carlo algorithm \algname{ForestCFCM}. Based on sampling forests, \algname{ForestCFCM} is pleasingly parallelizable while maintaining an approximation factor for solving CFCM. To further enhance efficiency and effectiveness, we develop another greedy algorithm \algname{SchurCFCM}. Based on the estimation of Schur complement, \algname{SchurCFCM} accelerates sampling spanning forests and enhances the quality of CFCM solutions, while still preserving an approximation factor. Numerical experiments validate the superiority of \algname{SchurCFCM} over both \algname{ForestCFCM} and the state-of-the-art method in terms of efficiency and effectiveness.

Our contributions of this work are summarized as follows.
\begin{itemize}
    \item We propose two Monte Carlo algorithms, \algname{ForestCFCM} and \algname{SchurCFCM}, for approximately maximizing CFCC \(\cfcc{S}\) under the cardinality constraint \(\abs{S}=k\). Both algorithms sample rooted spanning forests, with \algname{SchurCFCM} additionally leveraging Schur complement.
    \item Both of our algorithms achieve an approximation factor of \(1-\frac{k}{k-1}\frac{1}{\rme}-\epsilon\) for \(0<\epsilon<1\). Meanwhile, \algname{SchurCFCM} exhibits a nearly linear time complexity.
    \item We conduct extensive experiments on real-world graphs. The results demonstrate that both of our proposed algorithms are up to \(370\times\) faster than the state-of-the-art method, while maintaining superior effectiveness.
\end{itemize}

\section{Preliminaries}

\begin{comment}
In this section, we briefly introduce some notations as well as some basic concepts on graphs, resistance distances, and random walks. We also discuss the design and performance of the state-of-the-art method.
\end{comment}

\subsection{Notations}

We use normal lowercase letters like \(a,b,c\) to denote real scalars, use bold lowercase letters like \(\veca,\vecb,\vecc\) to denote vectors, and use bold uppercase letters like \(\mata,\matb,\matc\) to denote matrices.

To represent specific elements in matrices, we use \(\matentry{\mata}{i}{j}\) to denote entry \((i,j)\) of matrix \(\mata\).
% We also use \(\mata_{[i,:]}\) and \(\mata_{[:,j]}\) to respectively denote the \(\myord{i}\) row and the \(\myord{j}\) column of matrix \(\mata\).
Moreover, we write sets in subscripts to denote subvectors and submatrices. For the case of subvectors, \(\veca_{-S}\) represents the subvector of \(\veca\) obtained by removing elements with indices in set \(S\). For the case of submatrices, \(\matentry{\mata}{U}{T}\) denotes the submatrix of \(\mata\) with row indices in \(U\) and column indices in \(T\), and \(\mata_{-S}\) represents the submatrix of \(\mata\) obtained by removing elements with row indices or column indices in \(S\). \inlcomm{Note that the subscript takes precedence over the superscript, thus \(\mata_{-S}^{-1}\) denotes the inverse of \(\mata_{-S}\) rather than the submatrix of \(\mata^{-1}\).}
Finally, we use \(\vece_i\) to denote the \(\myord{i}\) standard basis vector, and \(\vecone\) to denote a vector with all elements being \(1\). \tabref{tab:notation} lists the frequently used notations throughout this paper.

\begin{table}[htbp]
    \centering
    \caption{Frequently used notations.}
    \label{tab:notation}
    \begin{tabularx}{\linewidth}{rX}
        \toprule
        \textbf{Notation}           & \textbf{Description}                                                                                                           \\
        \midrule
        \(\gr=(V,E)\)               & A graph with node set \(V\) and edge set \(E\).                                                                                \\
        \(n,m\)                     & The number of nodes and edges in \(\gr\).                                                                                      \\
        \(\tau\)                    & The diameter of \(\gr\).                                                                                                       \\
        \(\dmax\)                   & The maximum degree of nodes in \(\gr\).                                                                                        \\
        \(\dmax(S)\)                & The maximum degree of nodes in the subgraph obtained from \(\gr\) by removing nodes in set \(S\) and and their incident edges. \\
        % \(\matp\)                   & The transition matrix of random walks on \(\gr\).                                                                                                                             \\
        \(\parent{u},\rootnode{u}\) & The parent node and root node of \(u\) in a spanning forest.                                                                   \\
        \(\forestfreq{u}{S}{a}{b}\) & The number of sampled forests where path from \(u\) to root set \(S\) traverses edge \(\edge{a}{b}\) from \(a\) to \(b\).      \\
        \(\rootnum{u}{t}\)          & The number of spanning forests where \(u\) is rooted at \(t\).                                                                 \\
        \(\voltage{u}{v}{S}\)       & The voltage at \(v\) with unit current flowing from \(u\) to \(S\).                                                            \\
        \(\meanvolt{u}{v}{S}\)      & The unbiased estimator of \(\vece_u^\top\invsublap\vece_v\).                                                                   \\
        \(\meanvolts{\vecw}{u}{S}\) & The unbiased estimator of \(\vecw^\top\invsublap\vece_u\).                                                                     \\
        \(\schcompl{\matm}{T}\)     & The Schur complement of matrix \(\matm\) onto subset \(T\).                                                                    \\
        \bottomrule
    \end{tabularx}
\end{table}

Since we prove the approximation guarantee of our algorithms in Sections~\ref{subsec:alg-forestcfcm} and~\ref{subsec:alg-schurcfcm}, it is necessary to give the definition of approximate factor.
\begin{definition}[\(\epsilon\)-approximation]
    Let \(x\) and \(\tilde{x}\) be positive scalars,  and let \(\epsilon\) be an error parameter satisfying \(\epsilon\in(0,1)\).
    Then \(\tilde{x}\) is an \(\epsilon\)-approximation of \(x\) if \((1-\epsilon)x\leq \tilde{x}\leq(1+\epsilon)x\) holds, which we denote as \(\tilde{x}\approx_{\epsilon}x\).
\end{definition}

\subsection{Graphs and Laplacian Matrices}\label{subsec:graph-lap}

We use \(\gr=(V,E)\) to denote connected undirected graph with \(n=\abs{V}\) nodes and \(m=\abs{E}\) edges, where \(V\) and \(E\) denote, respectively, the node set and edge set of \(\gr\). We use \(e=\edge{u}{v}\) to denote an edge \(e\) linking node \(u\) and node \(v\).

The adjacency matrix of \(\gr\) is denoted as \(\mata\in \rea^{n\times n}\): for two nodes $i\in V$ and \(j\in V\), \(\matentry{\mata}{i}{j}=\matentry{\mata}{j}{i}=1\) if \(i\) and \(j\) are adjacent, and \(\matentry{\mata}{i}{j}=\matentry{\mata}{j}{i}=0\) otherwise. The degree vector of \(\gr\) is defined as \(\vecd=\mata\vecone=\spar{d_1,d_2,\dots,d_n}^\top\), where \(d_i\) represents the degree of node \(i\). If we denote the degree diagonal matrix as \(\matd=\diag{d_1,d_2,\cdots,d_n}\), then the Laplacian matrix \(\lap\) of \(\gr\) is defined as \(\lap=\matd-\mata\).
% It is easy to prove that \(\lap\) is positive semi-definite. In addition, all its eigenvalues are positive except for one unique zero eigenvalue. If we denote eigenvalues and corresponding eigenvectors of \(\lap\in\rea^{n\times n}\) as \(0=\lambda_1<\lambda_2\leq\lambda_3\leq\dots\leq\lambda_n\) and \(\vecv_1,\vecv_2,\dots,\vecv_n\) respectively, then \(\lap\) can be rewritten as \(\lap=\sum_{i=1}^n\lambda_i\vecv_i\vecv_i^\top\).
Since \(\lap\) is not invertible due to its null space \(\vecone\), we turn to use its pseudoinverse \(\lap^{\dagger}=\bigparinv{\lap+\frac{1}{n}\vecone\vecone^\top}-\frac{1}{n}\vecone\vecone^\top\). \(\lap^{\dagger}\) appears in many quantities related to random walks, such as the Kemeny constant~\cite{Hu14}. Moreover, It is easy to verify that Laplacian matrix \(\lap\) and its submatrices \(\sublap\) are Symmetric Diagonally Dominant (SDD).

\subsection{Spanning Forests and Random Walks}

For a graph \(\gr=(V,E)\), a spanning subgraph of \(\gr\) retains all nodes from \(V\) while including only a subset of edges from \(E\). A spanning forest is a spanning subgraph of \(\gr\), whose connected components are trees or isolated nodes. A rooted spanning forest of \(\gr\) is a spanning forest of \(\gr\) where a particular node in each tree is designated as its root. The set of all root nodes forms the root set of this spanning forest.

Many studies focus on methods for uniformly sampling rooted spanning forests, including Wilson's algorithm~\cite{Wi96} based on loop-erased random walks. For a connected graph, a classical random walk is defined by its transition matrix \(\matp\). At each step, if the walker is at node \(i\), it moves to an adjacent node \(j\) with equal probability \(\matentry{\matp}{i}{j}\). It follows that \(\matp=\matd^{-1}\mata\). To generate a spanning forest with root set \(S\), the loop-erased random walk is simulated iteratively. Initially \(\forest=S\), when the random walker visits a node in \(\forest\), the loop-erased path from the source node to this node is added into \(\forest\). Notably, it has been proved in~\cite{Wi96} that the distribution of forests sampled by Wilson's algorithm is independent of the order of source nodes.

\subsection{Resistance Distance and Relevant Centrality}

For an arbitrary undirected graph \(\gr=(V,E)\), we  define its corresponding electrical network by treating edges as unit resistors and nodes as junctions between resistors~\cite{DoSn84}. \inlcomm{The resistance value of all associated edges are \(1\).} For graph \(\gr\), if we denote \(\voltage{i}{u}{j}\) as the voltage of node \(u\) when a unit current enters at node \(i\) and leaves at node \(j\), then the resistance distance \(\resist{i}{j}\triangleq\voltage{i}{i}{j}-\voltage{i}{j}{j}\) is defined as the effective resistance between \(i\) and \(j\) in the corresponding electrical network~\cite{KlRa93}.

For a connected graph \(\gr\), the resistance distance \(\resist{i}{j}\) between two nodes \(i\) and \(j\) can be expressed in terms of \(\lap^{\dagger}\)~\cite{KlRa93,FoPiReSa07} and the inverse for submatrices of \(\lap\)~\cite{GhBoSa08} as:
\begin{gather}
    \resist{i}{j}=\matentry{\lap^{\dagger}}{i}{i}+\matentry{\lap^{\dagger}}{j}{j}-2\matentry{\lap^{\dagger}}{i}{j},\label{eq:resist-lap}\\
    \resist{i}{j}=\matentry{\bigspar{\invsublap[i]}}{j}{j}=\matentry{\bigspar{\invsublap[j]}}{i}{i}.\label{eq:resist-sublap-diag}
\end{gather}
% \begin{equation}\label{eq:resist-lap}
%     \resist{i}{j}=\matentry{\lap^{\dagger}}{i}{i}+\matentry{\lap^{\dagger}}{j}{j}-2\matentry{\lap^{\dagger}}{i}{j},
% \end{equation}
% \begin{equation}\label{eq:resist-sublap-diag}
%     \resist{i}{j}=\matentry{\bigspar{\invsublap[i]}}{j}{j}=\matentry{\bigspar{\invsublap[j]}}{i}{i}.
% \end{equation}
Since the resistance distance exhibits properties of a metric, we can use it to represent the proximity between node pairs, and measure the importance of each node by defining relevant centrality, such as the current flow closeness centrality (CFCC)~\cite{BrFl05}. For an \(n\)-node connected graph, the CFCC of a node \(u\) is defined as \(\cfcc{u}\triangleq n/\spar{\sum_{v\in V}\resist{u}{v}}\). According to~\eqref{eq:resist-lap}, we can easily represent \(\cfcc{u}\) in terms of the diagonal elements of \(\lap^{\dagger}\) as \(\cfcc{u}={n}/\bigspar{\bigtrace{\lap^{\dagger}}+n\matentry{\lap^{\dagger}}{u}{u}}\).
% \begin{equation}
%     \cfcc{u}=\frac{n}{\bigtrace{\lap^{\dagger}}+n\matentry{\lap^{\dagger}}{u}{u}}.
% \end{equation}

\subsection{Group Centrality and Its Maximization}

To measure the importance of node groups, Li~\textit{et al.}~\cite{LiPeShYiZh19} extended the concept of CFCC to multiple nodes. In defining CFCC for a node group \(S\), all nodes in \(S\) are assumed to be grounded, with their voltages being consistently at \(0\). Therefore, we denote the voltage of node \(v\) when a unit current enters at node \(u\) and leaves at an arbitrary node in \(S\) as \(\voltage{u}{v}{S}\). According to~\cite{ClPo11}, the resistance distance \(\resist{u}{S}\) between an individual node \(u\) and a grounded node group \(S\) is defined as \(\voltage{u}{u}{S}=\matentry{\bigspar{\invsublap}}{u}{u}\). Therefore, for an \(n\)-node graph, the CFCC of a node group \(S\) can be defined as~\cite{LiPeShYiZh19}:
\begin{equation}\label{eq:def-cfgcc}
    \cfcc{S}\triangleq\frac{n}{\sum_{u\in V}\resist{u}{S}}=\frac{n}{\bigtrace{\invsublap}}.
\end{equation}
Equation~\eqref{eq:def-cfgcc} illustrates that a smaller sum of \(\resist{u}{S}\) indicates greater accessibility for a node \(u\in\complset\) to nodes in \(S\), suggesting that \(S\) is more significant in the network. This connection between accessibility and importance forms the basis for terming this measure as group centrality. Therefore, the problem of finding the most important node group with cardinality constraint is naturally given in~\cite{LiPeShYiZh19}:
\begin{definition}[Current Flow Closeness Maximization, CFCM]
    For an \(n\)-node graph \(\gr=(V,E)\) and an integer \(k\ll n\), the goal is to find a node group \(S^*\) of size \(k\) that maximizes its CFCC. In other words, \(S^*\triangleq\argmax\limits_{S\subset V,\abs{S}=k}\cfcc{S}\).
    % \begin{equation*}
    %     S^*\triangleq\argmax_{S\subset V,\abs{S}=k}\cfcc{S}.
    % \end{equation*}
\end{definition}

\begin{comment}
\subsection{Monotonicity and Supermodularity}

Subsequently, we give the definitions of monotone and supermodular set functions. For simplicity, we denote \(S\cup\setof{u}\) as \(S+u\).
\begin{definition}[Monotonicity]
    The set function \(f:2^{V}\to\rea^+\) is monotone if and only if for any non-empty set \(X,Y\) that satisfies \(X\subset Y\subset V\), the inequality \(f(X)\ge f(Y)\) holds.
\end{definition}
\begin{definition}[Supermodularity]
    The set function \(f:2^{V}\to\rea^+\) is supermodular if and only if for any non-empty set \(X,Y\) that satisfies \(\forall X\subset Y\subset V, u\in\complset[Y]\), the inequality \(f(X)-f(X+u)\ge f(Y)-f(Y+u)\) holds.
\end{definition}
\end{comment}

\subsection{State-of-the-art Method}\label{subsec:exist-method}

% In this section, we present an overview of the greedy algorithm \algname{ApproxGreedy} proposed by Li~\textit{et al.}~\cite{LiPeShYiZh19} for solving CFCM, which is based on a fast Laplacian solver. After describing the design of \algname{ApproxGreedy}, we also discuss its limitation.

\textbf{Algorithm Design.}
As a combinatorial optimization problem, CFCM is proved in~\cite{LiPeShYiZh19} to be NP-hard. Therefore, designing a polynomial-time algorithm to solve CFCM is nearly impossible. However, the reciprocal of CFCC is also proved in~\cite{LiPeShYiZh19} to be monotone and supermodular. These findings indicate that a naive greedy strategy yields a solution with an approximation factor of \(1-\frac{k}{k-1}\frac{1}{\mathrm{e}}\)~\cite{NeWoFi78}. Set \(S\) is initially empty, then \(k\) nodes from \(\complset\) are added to \(S\) iteratively. For the first iteration, the task is to find the node \(u\in V\) with minimum sum of resistance distances, which is expressed as
\begin{equation}\label{eq:def-gain-first}
    \sum_{v\in V}\resist{u}{v}=\bigtrace{\lap^{\dagger}}+n\matentry{\lap^{\dagger}}{u}{u}.
\end{equation}
For subsequent iterations, the node \(u\in\complset\) with maximum marginal gain \(\gain{u}{S}\) is selected, which is expressed as
\begin{equation}\label{eq:def-gain-subseq}
    \small
    \gain{u}{S}\triangleq\bigtrace{\invsublap}-\bigtrace{\invsublap[(S+u)]}=\frac{\matentry{\bigspar{\sqinvsublap}}{u}{u}}{\matentry{\bigspar{\invsublap}}{u}{u}}\ (S\neq\varnothing).
\end{equation}
Equations~\eqref{eq:def-gain-first} and~\eqref{eq:def-gain-subseq} indicate that computing the marginal gain involves calculating diagonal elements of \(\lap^{\dagger}\) for the first iteration and diagonal elements of \(\invsublap\) and \(\sqinvsublap\) for subsequent iterations. Leveraging the Johnson-Lindenstrauss lemma~\cite{JoLi84}, the state-of-the-art method \algname{ApproxGreedy} transforms estimating diagonal elements of matrix inverse into solving linear equations related to the Laplacian matrix. Rather than directly solving linear equations, \algname{ApproxGreedy} introduces a nearly linear-time Laplacian solver~\cite{KySa16}.

\textbf{Performance Discussion.}
The key limitation of \algname{ApproxGreedy} stems from its reliance on Laplacian solver. Each iteration requires solving \(\normo{\epsilon^{-1}\log n}\) linear equations, yielding a time complexity of \(\tildeo{k\epsilon^{-3}m}\) dominated by the number \(m\) of edges. This creates two fundamental bottlenecks. First, for dense graphs where \(m=\normo{n^2}\), the quadratic scaling becomes prohibitive. Second, even on sparse graphs, \algname{ApproxGreedy} depends on highly optimized implementations of Laplacian solver. This dependency limits deployment in environments where computing resource is constrained or language flexibility is critical.

Our proposed algorithms address these limitations by replacing Laplacian solvers with other techniques. Through spanning forest sampling and improvements from Schur complement, the time complexity of our algorithms is nearly linear in the node number \(n\). These approaches enable better scalability to denser graphs, as demonstrated in \secref{sec:num-exp}.

\begin{comment}
According to~\cite{LiPeShYiZh19}, the time complexity of \algname{ApproxGreedy} is reduced to \(\tilde{O}\normspar{k\epsilon^{-3}m}\), where \(m\) denotes the number of edges and \(\tilde{O}\spar{\cdot}\) notation omits poly-logarithmic factors. Experimental results from~\cite{LiPeShYiZh19} indicate that \algname{ApproxGreedy} can handle networks with more than 2 million edges. Nevertheless, the same study demonstrates that the longest running time of \algname{ApproxGreedy} exceeds 7 hours, rendering it impractical for many real-world scenarios. Furthermore, the time complexity of \algname{ApproxGreedy} is dominated by the number of edges rather than the number of nodes, which limits its performance on graphs with numerous edges. In conclusion, \algname{ApproxGreedy} falls short of meeting the requirements of analyzing large-scale graphs, and should be substituted by an algorithm with time complexity primarily dependent on the number of nodes.
\end{comment}

\section{Algorithm Based on Forest Sampling}

In this section, we propose a greedy algorithm to approximately solve CFCM through sampling rooted spanning forests, addressing the challenge of estimating the marginal gain defined in~\eqref{eq:def-gain-first} and~\eqref{eq:def-gain-subseq}.
% We first demonstrate that, under the model of electrical networks, there exists an intimate relationship between \(\invsublap\) and the voltage of nodes in random spanning forests. Moreover, we show that diagonal elements of \(\lap^{\dagger}\) can also be expressed in terms of \(\invsublap\), leading to unbiased estimators for these quantities.
This section proceeds as follows.
We first derive unbiased estimators for entries of \(\invsublap\) via connections between spanning forests and electrical network analysis (\lemref{lem:estim-sublap}). Next, we approximate diagonal elements of \(\sqinvsublap\) by reducing its dimension (\lemref{lem:jl}). Additionally, we reformulate diagonal elements of \(\lap^{\dagger}\) in terms of \(\invsublap\) to overcome weak diagonal dominance (\lemref{lem:sublap-pseudolap}). An adaptive sampling strategy using empirical Bernstein inequality is then introduced to optimize efficiency. Finally, these components culminate in the algorithm \algname{ForestCFCM}, which achieves nearly linear complexity and rigorous approximation guarantees (\thmref{thm:perf-forestcfcm}).

\subsection{Estimation of Laplacian Submatrix Inverse}\label{subsec:estim-sublap}

For an undirected graph \(\gr\), let \(\treemark\) denote the total number of spanning trees. Given two nodes \(u,v\) and an edge \(\edge{a}{b}\), we denote \(\treenum{u}{v}{a}{b}\) as the number of spanning trees where the unique path from \(u\) to \(v\) traverses edge \(\edge{a}{b}\) from \(a\) to \(b\). According to~\cite{Sh87}, we have

\begin{lemma}\label{lem:current-spantree}
    In the corresponding electrical network of a graph \(\gr\), suppose a unit current flows from \(u\) to \(v\). The current through \(\edge{a}{b}\) is then given by \(\frac{1}{\treemark}\bigspar{\treenum{u}{v}{a}{b}-\treenum{u}{v}{b}{a}}\).
\end{lemma}

% \lemref{lem:current-spantree} implies that for a unit current with specified source and target nodes, the current through any edge can be approximated by sampling random spanning trees and calculating the mean value of the directed traversal probability for this edge.
Since CFCC is defined for node groups, we subsequently extend \lemref{lem:current-spantree} to the case of multiple target nodes. Similarly, we denote \(\forestmark(S)\) as the total number of spanning forests with root set \(S\), and let \(\forestnum{u}{S}{a}{b}\) denote the number of spanning forests where the unique path from \(u\) to an arbitrary node in \(S\) traverses edge \(\edge{a}{b}\) in the direction from \(a\) to \(b\).

\begin{lemma}\label{lem:current-spanforest}
    Let \(\gr\) be a graph with a node group \(S\). For a unit current flowing from \(u\) to an arbitrary node in the grounded node group \(S\), the current through \(\edge{a}{b}\) is given by \(\frac{1}{\forestmark(S)}\bigspar{\forestnum{u}{S}{a}{b}-\forestnum{u}{S}{b}{a}}\).
\end{lemma}
\begin{IEEEproof}
    To extend \lemref{lem:current-spantree} to multiple targets, we map spanning forests in \(\gr\) to spanning trees in an augmented graph \(\gr^* = (V \cup \{r^*\}, E \cup \{\edge{r^*}{s} | s \in S\})\). Each spanning forest of \(\gr\) rooted at \(S\) bijectively corresponds to a spanning tree of \(\gr^*\) rooted at \(r^*\). Applying \lemref{lem:current-spantree} to \(\gr^*\) with unit current from \(u\) to \(r^*\), the current through edge \(\edge{a}{b}\) in \(E\) equals \(\frac{1}{N}(\treenum{u}{r^*}{a}{b} - \treenum{u}{r^*}{b}{a})\). Translating via the forest-tree correspondence yields \(\frac{1}{\forestmark(S)}(\forestnum{u}{S}{a}{b} - \forestnum{u}{S}{b}{a})\) for \(\gr\), completing the proof.
\end{IEEEproof}

Given a graph \(\gr\) and a unit current with source node \(u\) and grounded target node group \(S\), \lemref{lem:current-spanforest} enables us to approximate the current through any edge of \(\gr\). Leveraging this, we can estimate the voltage of any node in \(\gr\), yielding an unbiased estimator of \(\vece_u^\top\invsublap\vece_v\).

\begin{lemma}\label{lem:estim-sublap}
    Let \(\gr\) be a graph with a node group \(S\). Suppose we sample \(\tildeforestmark(S)\) spanning forests with root set \(S\), and let \(\forestfreq{u}{S}{a}{b}\) denote the number of sampled forests where the unique path from \(u\) to an arbitrary node in \(S\) traverses edge \(\edge{a}{b}\) from \(a\) to \(b\). Then the quantity
    \begin{equation*}
        \meanvolt{u}{v}{S}\triangleq\frac{1}{\tildeforestmark(S)}\sum_{\edge{a}{b}\in\pathway{v}{S}}\bigspar{\forestfreq{u}{S}{a}{b}-\forestfreq{u}{S}{b}{a}}
    \end{equation*}
    is an unbiased estimator of \(\vece_u^\top\invsublap\vece_v\), where \(\pathway{v}{S}\) denotes edges on paths from \(v\) to an arbitrary node in \(S\).
\end{lemma}
\begin{IEEEproof}
    According to~\cite{ClPo11}, the voltage \(\voltage{u}{v}{S}\) is equal to \(\matentry{\bigspar{\invsublap}}{u}{v}\), which also equals the sum of currents along any path from \(v\) to \(S\) by Kirchhoff's law. Recall from \lemref{lem:current-spanforest} that \(\frac{1}{\tildeforestmark(S)}\bigspar{\forestfreq{u}{S}{a}{b}-\forestfreq{u}{S}{b}{a}}\) unbiasedly estimates the current in edge \(\edge{a}{b}\), summing these estimators over \(\pathway{v}{S}\) and applying linearity of expectation yields \(\mean{\meanvolt{u}{v}{S}}=\voltage{u}{v}{S}\).
\end{IEEEproof}

\subsection{Estimation of Diagonal Elements for Squared Laplacian Submatrix Inverse}\label{subsec:estim-sqsublap}

Despite the relationship between \(\invsublap\) and random spanning forests, there is no obvious physical meaning of \(\sqinvsublap\). Therefore, we transform estimating the diagonal element \(\vece_u^\top\sqinvsublap\vece_u\) into approximating the squared \(2\)-norm \(\Abs{\invsublap\vece_u}^2\). However, the time of directly computing one squared \(2\)-norm in \(\rea^n\) is \(\Omega(n)\). To accelerate this evaluation, we introduce the Johnson-Lindenstrauss Lemma (JL Lemma)~\cite{JoLi84}.

\begin{lemma}[JL Lemma~\cite{JoLi84}]\label{lem:jl}
    Given \(n\) fixed vectors \(\vecv_1,\vecv_2,\dots,\vecv_n\in\rea^d\) and \(0<\epsilon<1\), let \(\matq\in\rea^{k\times d}\) be a matrix with each entry equal to \(\pm k^{-\nicehalf}\) with equal probability. If \(k\geq24\epsilon^{-2}\log n\), then \(\Abs{\vecv_i-\vecv_j}^2\approx_{\epsilon}\Abs{\matq\vecv_i-\matq\vecv_j}^2\) holds with probability at least \(1-\nicefrac{1}{n}\) for all pairs \(i,j\leq n\).
    % \begin{equation*}
    %     \Abs{\vecv_i-\vecv_j}^2\approx_{\epsilon}\Abs{\matq\vecv_i-\matq\vecv_j}^2
    % \end{equation*}
\end{lemma}

\lemref{lem:jl} indicates that the pairwise distances of vectors \(\vecv_i\in\rea^d\) are almost preserved if we project them into a lower-dimensional space spanned by \(\normo{\log n}\) random vectors. Therefore, we can use this lemma to reduce the dimension of \(\invsublap\vece_u\) with bounded error. Concretely, we denote \(\matw\in\rea^{w\times n}\) as random matrix where \(w\geq24\epsilon^{-2}\log n\), then we have
\begin{equation}\label{eq:sqsublap-jl}
    \vece_u^\top\sqinvsublap\vece_u=\Abs{\invsublap\vece_u}^2\approx_{\epsilon}\Abs{\matw\invsublap\vece_u}^2.
\end{equation}
Incorporating~\eqref{eq:sqsublap-jl} into \lemref{lem:estim-sublap}, we transform estimating \(\vece_u^\top\sqinvsublap\vece_u\) into solving linear equations \(\vecw_i^\top\invsublap\) for \(1\leq i\leq w\). From this, we can further derive the unbiased estimator of \(\vece_u^\top\sqinvsublap\vece_u\). Specifically, the unbiased estimator for \(\vecw_i^\top\invsublap\vece_u\) is defined as \(\meanvolts{\vecw_i}{u}{S}\triangleq\sum_{v\in\complset}w_{iv}\meanvolt{v}{u}{S}\), which is represented as
\begin{equation*}\label{eq:estim-sqsublap}
    \meanvolts{\vecw_i}{u}{S}=\frac{1}{\tildeforestmark(S)}\sum_{\edge{a}{b}\in\pathway{u}{S}}\sum_{v\in\complset}w_{iv}\bigspar{\forestfreq{v}{S}{a}{b}-\forestfreq{v}{S}{b}{a}}.
\end{equation*}

\subsection{Estimation of Diagonal Elements for Laplacian Pseudoinverse}\label{subsec:estim-pseudolap}

% After obtaining the unbiased estimator for elements in \(\invsublap\) and diagonal elements in \(\sqinvsublap\), the efficient diagonal estimation of \(\lap^{\dagger}\) remains a challenge.
The challenge of estimating \(\lap^{\dagger}\) arises from the weak diagonal dominance of \(\lap\), which reduces efficiency and effectiveness of approximation algorithms. To address this issue, we represent \(\vece_u^\top\lap^{\dagger}\vece_u\) in terms of \(\invsublap\)~\cite{BoRaZh11}, which leads to better diagonal dominance and another unbiased estimator.

\begin{lemma}[\cite{BoRaZh11}]\label{lem:sublap-pseudolap}
    When \(S=\setof{s}\), we have
    \begin{equation*}
        \small
        \vece_u^\top\lap^{\dagger}\vece_u=
        \begin{cases}
            \vece_u^\top\invsublap\vece_u-\frac{2}{n}\vecone^\top\invsublap\vece_u+\frac{1}{n^2}\vecone^\top\invsublap\vecone & u\neq s \\
            \frac{1}{n^2}\vecone^\top\invsublap\vecone                                                                        & u=s
        \end{cases}.
    \end{equation*}
\end{lemma}

\lemref{lem:sublap-pseudolap} establishes the connection between diagonal elements of \(\lap^{\dagger}\) and elements of \(\invsublap\), where \(S\) contains only one node \(s\). Combining it with \lemref{lem:estim-sublap}, we can define the unbiased estimator for \(\vece_u^\top\lap^{\dagger}\vece_u\) as linear combinations of \(\meanvolt{u}{v}{S}\). For instance, when \(S\) only contains one node \(s\), the unbiased estimator for \(\vecone^\top\invsublap\vece_u\) is defined as \(\meanvolts{\vecone}{u}{S}\triangleq\sum_{v\in\complset}\meanvolt{v}{u}{S}\), leading to the expression:
\begin{equation}\label{eq:estim-sum}
    \meanvolts{\vecone}{u}{S}=\frac{1}{\tildeforestmark(S)}\sum_{\edge{a}{b}\in\pathway{u}{S}}\sum_{v\in\complset}\bigspar{\forestfreq{v}{S}{a}{b}-\forestfreq{v}{S}{b}{a}}.
\end{equation}
Regarding the estimation of \(\frac{1}{n^2}\vecone^\top\invsublap\vecone\), we observe that this term appears in \(\matentry{\lap^{\dagger}}{u}{u}\) for every node \(u\in V\). Recall from~\eqref{eq:def-gain-first} that we only need to identify the node \(u\in V\) with minimum \(\matentry{\lap^{\dagger}}{u}{u}\), we can omit the computation of this term without introducing any error.

\subsection{Adaptive Forest Sampling}

After deriving an unbiased estimation of the marginal gain, we need to determine the upper bound of sample size to achieve an approximation factor for solving CFCM. \lemref{lem:perf-forestdelta} uses Hoeffding's inequality to derive an upper bound. However, this inequality does not consider the variance of random variables, providing a loose theoretical bound. We thus resort to the empirical Bernstein inequality~\cite{HuSeTa07}:

\begin{lemma}\label{lem:bernstein}
    Let \(X_1,X_2,\dots,X_n\) be \(n\) real-valued i.i.d. random variables that satisfy \(0\leq X_i\leq X_{\mathrm{sup}}\). If we denote \(\bar{X}\) and \(X_{\mathrm{var}}\) as the empirical mean and the empirical variance of \(X_i\), then we have
    \begin{equation*}
        \Pr\spar{\abs{\bar{X}-\mathbb{E}[\bar{X}]}\geq \bernfunc{n,X_{\mathrm{var}},X_{\mathrm{sup}},\delta}}\leq\delta,
    \end{equation*}
    where \(\bernfunc{n,X_{\mathrm{var}},X_{\mathrm{sup}},\delta}\triangleq\sqrt{\frac{2X_{\mathrm{var}}\log\spar{\nicefrac{3}{\delta}}}{n}}+\frac{3X_{\mathrm{sup}}\log\spar{\nicefrac{3}{\delta}}}{n}\).
    % \begin{equation*}
    %     \bernfunc{n,X_{\mathrm{var}},X_{\mathrm{sup}},\delta}=\sqrt{\frac{2X_{\mathrm{var}}\log\spar{\nicefrac{3}{\delta}}}{n}}+\frac{3X_{\mathrm{sup}}\log\spar{\nicefrac{3}{\delta}}}{n}.
    % \end{equation*}
\end{lemma}

\lemref{lem:bernstein} differs from Hoeffding's inequality in that it incorporates the empirical variance of random variables. While the empirical variance remains unknown a priori, it can be efficiently maintained throughout the sampling process. We introduce the empirical Bernstein inequality while retaining the Hoeffding bound to preserve the theoretical guarantee. Meanwhile, if the empirical error of estimators provided by \lemref{lem:bernstein} falls below the threshold parameter, we terminate sampling rooted spanning forests. Notably, the approximation factor provided in \thmref{thm:perf-forestcfcm} remains unaffected by applying this adaptive strategy.

\subsection{Algorithm Design and Analyses}\label{subsec:alg-forestcfcm}

Based on the above analyses, we propose our first approximation algorithm \algname{ForestCFCM} for solving CFCM, which is depicted in \algref{alg:forestcfcm}. In each subsequent iteration of \algname{ForestCFCM}, it invokes \algname{ForestDelta}, which is depicted in \algref{alg:forestdelta}. Both of them sample random spanning forests through \algname{RandomForest}, which is depicted in \algref{alg:randomforest}.

\begin{algorithm}
    \caption{\algname{RandomForest}\((\gr,S)\)}
    \label{alg:randomforest}
    \begin{small}
        \Input{
            \(\gr=(V,E)\): an undirected graph;
            \(S\subset V\): the set of root nodes
        }
        \Output{
            \(\parent{u}\): the parent node of \(u\in\complset\) in the random spanning forest \(\rsf\) of \(\gr\);
            \(\dfslist\): the list of nodes visited in reverse DFS order on \(\rsf\)
        }
        \(\parent{u}\gets\) arbitrary value, \(\textnormal{InForest}_{u}\gets\textnormal{false}\) for \(u\in\complset\)\;
        \(\textnormal{InForest}_{s}\gets\textnormal{true}\) for \(s\in S\), \(\dfslist\gets\) an empty list\;
        \ForEach{\(u\in\complset\)}{
            \(i\gets u\)\;\label{line:lewalk-begin}
            \While{\(\textnormal{InForest}_{i}=\textnormal{false}\)}{
                \(\parent{i}\gets\) a randomly selected neighbor of \(i\)\;
                \(i\gets\parent{i}\)\;\label{line:lewalk-end}
            }
            \(i\gets u\), \(\textnormal{chain}\gets\) an empty list\;\label{line:loop-erase-begin}
            \While{\(\textnormal{InForest}_{i}=\textnormal{false}\)}{
                \(\textnormal{InForest}_{i}\gets\textnormal{true}\)\;
                Add \(i\) to the end of \(\textnormal{chain}\)\;
                \(i\gets\parent{i}\)\;\label{line:loop-erase-end}
            }
            Reversely add elements in \(\textnormal{chain}\) to the end of \(\dfslist\)\;\label{line:dfs-begin}
        }
        Reverse the node order in \(\dfslist\)\;\label{line:dfs-end}
        \Return{\(\setof{\parent{u}|u\in\complset}\), \(\dfslist\)}
    \end{small}
\end{algorithm}
\subsubsection{\algname{RandomForest}}
Based on Wilson's algorithm~\cite{Wi96}, \algref{alg:randomforest} iteratively simulates loop-erased random walks to sample spanning forests with root set \(S\). Starting from a node, each walk comprises two phases: simulation of a random walk (\linerangeref{line:lewalk-begin}{line:lewalk-end}) and erasure of loops within the walk path (\linerangeref{line:loop-erase-begin}{line:loop-erase-end}). During loop erasure, nodes are accessed in the order of a chain in the spanning forest.
Different from Wilson's algorithm, \algref{alg:randomforest} properly maintains the accessed order of nodes (\linerangeref{line:dfs-begin}{line:dfs-end}), resulting in a reverse DFS order \(\dfslist\) on the spanning forest. This maintenance stems from the need of \algname{ForestDelta} and \algname{ForestCFCM} for both traversing from each node \(u\in\complset\) to its root. By iterating over \(\dfslist\), the maximum number of visits for each node can be reduced from \(\tau\) to \(1\), where \(\tau\) denotes the diameter of graph.
The time complexity of \algname{RandomForest} is given in \lemref{lem:perf-randomforest}.

\begin{lemma}\label{lem:perf-randomforest}
    For a graph with root set \(S\), the upper bound of time complexity for \algref{alg:randomforest} is \(\bigo{\bigtrace{\parinv{\mati-\matp_{-S}}}}\). For a real-world graph with \(n\) nodes, this form can be expressed as \(\tildeo{n}\), where \(\tildeo{\cdot}\) omits poly-logarithmic factors.
\end{lemma}
\begin{IEEEproof}
    The time complexity of \algname{RandomForest} is dominated by the total number of visits to nodes during loop-erased random walks. In the first iteration of walk starting from node \(i\), the expected number of visits to unvisited nodes is \(\matentry{\parinv{\mati-\matp_{-S}}}{i}{i}\)~\cite{ZhYaLi12}. According to Wilson's algorithm~\cite{Wi96}, the sampling distribution is invariant to the order of source nodes, allowing any strategy of selecting the first starting node. Summing over all nodes gives the upper bound \(\bigtrace{\parinv{\mati-\matp_{-S}}}\).

    For real-world graphs, \(\bigtrace{\parinv{\mati-\matp_{-S}}}\) is bounded by \(\kem + \absorb{s}\) for any \(s \in S\)~\cite{XiZh24}, where \(\kem\) denotes Kemeny's constant and \(\absorb{s}\) denotes absorbing centrality. Both of them scale nearly linearly with \(n\) in scale-free networks~\cite{XuShZhKaZh20,LiJuZh12}. Given real-world graphs exhibit scale-free~\cite{BaAl99} and small-world~\cite{WaSt98} properties, the complexity reduces to \(\tildeo{n}\).
\end{IEEEproof}

\begin{algorithm}
    \caption{\algname{ForestDelta}\((\gr,S,\epsilon)\)}
    \label{alg:forestdelta}
    \begin{small}
        \Input{
            \(\gr=(V,E)\): an undirected \(n\)-node graph;
            \(S\subset V\): the node subset;
            \(\epsilon\): an error parameter
        }
        \Output{
            \(\approxgain{u}{S}\): the margin for node \(u\in\complset\)
        }
        \(w\gets\ceil{24(\nicefrac{\epsilon}{7})^{-2}\log n}\), \(r\gets{2(\nicefrac{\epsilon}{15})^{-2}\tau^2\dmax^{2\tau+2}(S)\log(2n)}\)\;
        \(\forestfreq{u}{S}{a}{b}\gets 0\) for \(u\in\complset\) and \(\edge{a}{b}\in E\)\;
        Construct matrix \(\matw\in\rea^{w\times \abs{\complset}}\) by \lemref{lem:jl}\;
        \(\bfslist\gets\) the list of nodes visited in BFS order from \(S\)\;\label{line:forestdelta-bfs-begin}
        \For{\(r'=1,2,\dots,\ceil{\log_2r}\)}{
        \ForPar{\(i=1,2,\dots,2^{r'}\)}{\label{line:forestdelta-forest-begin}
        \(\setof{\parent{v}|v\in\complset},\dfslist\gets\)\algname{RandomForest}\((\gr,S)\)\;
        \ForEach{\(u\in\dfslist\)}{
            Update \(\sum_{v\in\complset}\matentry{\matw}{j}{v}\forestfreq{v}{S}{u}{\parent{u}}\) for \(j=1,2,\dots,w\)\;\label{line:forestdelta-count-begin}
            Update \(\forestfreq{u}{S}{u}{\parent{u}}\)\;\label{line:forestdelta-forest-end}\label{line:forestdelta-count-end}
        }
        }
        \ifthenelse{\isundefined{\IEEEauthorblockN}}
        {\(\maty\in\rea^{w\times\abs{\complset}}\gets\) arbitrary value\tcp*{Estimator of \(\matw\invsublap\)}}
        {\tcp{Estimator of \(\matw\invsublap\)}\(\maty\in\rea^{w\times\abs{\complset}}\gets\) arbitrary value}

        \ForEach{\(u\in\bfslist\)}{
            \(\matentry{\maty}{j}{u}\gets\meanvolts{\vece_j^\top\matw}{u}{S}\) for \(j=1,2,\dots,w\)\;
            \(z_u\gets\meanvolt{u}{u}{S}\)\tcp*{Estimator of \(\matentry{\bigspar{\invsublap}}{u}{u}\)}\label{line:forestdelta-bfs-end}
        }
        \(\approxgain{u}{S}\gets z_u^{-1}\Abs{\maty\vece_u}^2\) for \(u\in\complset\)\;
        Compute error bound \(\epsilon'_u\) for \(\approxgain{u}{S}\) by \lemref{lem:bernstein}\;\label{line:forestdelta-bern-begin}
        \lIf{\({\epsilon'_u}\leq\epsilon\normspar{\approxgain{u}{S}-\epsilon'_u}\) for \(u\in V\setminus S\)}{break}\label{line:forestdelta-bern-end}
        }
        \Return{\(\setof{\approxgain{u}{S}=z_u^{-1}\Abs{\maty\vece_u}^2\big\vert u\in\complset}\)}
    \end{small}
\end{algorithm}
\subsubsection{\algname{ForestDelta}}
Given a graph \(\gr=(V,E)\), a node subset \(S\), and an error parameter \(\epsilon\), \algref{alg:forestdelta} estimates \(\gain{u}{S}=\frac{\vece_u^\top\sqinvsublap\vece_u}{\vece_u^\top\invsublap\vece_u}\) for \(u\in\complset\). To achieve this, \algname{ForestDelta} samples rooted spanning forests of \(\gr\) (\linerangeref{line:forestdelta-forest-begin}{line:forestdelta-forest-end}). To estimate the numerator and denominator of \(\gain{u}{S}\), \algname{ForestDelta} efficiently maintains counters of different sampled forests (\linerangeref{line:forestdelta-count-begin}{line:forestdelta-count-end}) via DFS order \(\dfslist\). Based on these counters, \algname{ForestDelta} computes the unbiased estimators described in Sections~\ref{subsec:estim-sublap} and~\ref{subsec:estim-sqsublap} along BFS order. To obtain its relative error bound, we employ Hoeffding's inequality.

\begin{lemma}[Hoeffding's inequality]\label{lem:hoeffding}
    Let \(x_1,x_2,\dots,x_n\) be \(n\) independent random variables that satisfy \(a\leq x_i\leq b\) for \(i=1,2,\dots,n\). Let \(x=\sum_{i=1}^{n}x_i\), then for any \(\epsilon>0\),
    \begin{equation*}
        \Pr\spar{\abs{x-\mean{x}}\geq\epsilon}\leq2\exp\setof{-\frac{2\epsilon^2}{n\spar{b-a}^2}}.
    \end{equation*}
\end{lemma}

Next, \lemref{lem:error-forestdelta} provides the relative error bound of \algname{ForestDelta}.
\begin{lemma}\label{lem:error-forestdelta}
    For an undirected graph \(\gr\) and the error parameter \(\epsilon\), if the sample size \(r\) satisfies
    \begin{equation}\label{eq:error-forestdelta}
        r\geq2(\nicefrac{\epsilon}{15})^{-2}\tau^2\dmax^{2\tau+2}(S)\log(2n),
    \end{equation}
    \algref{alg:forestdelta} returns \(\approxgain{u}{S}\approx_{\epsilon}\gain{u}{S}\) for \(u\in V\setminus S\) with probability of \(1-\nicefrac{1}{n}\).
\end{lemma}
\begin{IEEEproof}
    We bound the denominator \(\matentry{\bigspar{\invsublap}}{u}{u}\) and numerator \(\matentry{\bigspar{\sqinvsublap}}{u}{u}\) of \(\gain{u}{S}\) separately.

    For \(\matentry{\bigspar{\invsublap}}{u}{u}\), its lower bound \(d_u^{-1}\) follows from Neumann series expansion of \(d_u^{-1}\matentry{\bigparinv{\mati-\matp_{-S}}}{u}{u}\). Since the upper bound of \(\bigabs{\meanvolt{u}{u}{S}}\) is \(\tau\), applying Hoeffding's inequality with sample size with~\eqref{eq:error-forestdelta} yields:
    \begin{equation}\label{eq:estim-forest-denom}
        \meanvolt{u}{u}{S} \approx_{\epsilon/3} \matentry{\bigspar{\invsublap}}{u}{u}.
    \end{equation}

    For the numerator \(\matentry{\bigspar{\sqinvsublap}}{u}{u}=\Abs{\invsublap\vece_u}^2\), we first apply \lemref{lem:jl} for dimension reduction, which boils down to estimate terms of \(\matentry{\bigspar{\invsublap}}{u}{v}\). Similar to that of \(\matentry{\bigspar{\invsublap}}{u}{u}\), the lower bound \(\dmax^{-\tau-1}(S)\) for \(\matentry{\bigspar{\invsublap}}{u}{v}\) combined with Hoeffding's inequality and~\eqref{eq:error-forestdelta} gives:
    \begin{equation}\label{eq:estim-forest-numer}
        \meanvolt{u}{v}{S} \approx_{\epsilon/15} \matentry{\bigspar{\invsublap}}{u}{v}.
    \end{equation}

    Combining~\eqref{eq:estim-forest-denom} and~\eqref{eq:estim-forest-numer} with reduced dimension \(w \geq 24(\epsilon/7)^{-2}\log n\), we obtain
    \begin{equation*}
        \frac{|\approxgain{u}{S} - \gain{u}{S}|}{\gain{u}{S}} \leq \frac{(1+\epsilon/15)^2(1+\epsilon/7)}{1-\epsilon/3} - 1 \leq \epsilon
    \end{equation*}
    for any \(0 < \epsilon < 1\), completing the proof.
\end{IEEEproof}

Combining Lemmas~\ref{lem:perf-randomforest} and~\ref{lem:error-forestdelta}, we can easily derive the performance of \algname{ForestDelta} as \lemref{lem:perf-forestdelta}.
% Despite the nearly-linear time complexity of \algname{ForestDelta}, this bound remains relatively loose due to the inclusion of \lemref{lem:bernstein} (\linerangeref{line:forestdelta-bern-begin}{line:forestdelta-bern-end}).

\begin{lemma}\label{lem:perf-forestdelta}
    For a realistic graph \(\gr\) with \(n\) nodes, the time complexity of \algref{alg:forestdelta} is \(\tildeo{\epsilon^{-2}n\tau^2\dmax^{2\tau+2}(S)}\). \algref{alg:forestdelta} returns \(\approxgain{u}{S}\approx_{\epsilon}\gain{u}{S}\) for \(u\in V\setminus S\) with probability of \(1-\nicefrac{1}{n}\).
\end{lemma}

\begin{algorithm}
    \caption{\algname{ForestCFCM}\((\gr,k,\epsilon)\)}
    \label{alg:forestcfcm}
    \begin{small}
        \Input{
            \(\gr=(V,E)\): an undirected \(n\)-node graph;
            \(k\ll n\): the capacity of node set;
            \(\epsilon\): an error parameter
        }
        \Output{
            \(S_k\): A subset of \(V\) with \(\abs{S_k}=k\)
        }
        \(s\gets\argmax_{u\in V}d_u\), \(r\gets{18\epsilon^{-2}\tau^2d_s^2(1-\nicefrac{1}{n})^{-4}\log(2n)}\)\;
        \(\forestfreq{u}{\setof{s}}{a}{b}\gets 0\) for \(u\in\complset[\setof{s}]\) and \(\edge{a}{b}\in E\)\;\label{line:forestcfcm-approx-init-begin}
        \(\bfslist\gets\) the list of nodes visited in BFS order from \(s\)\;\label{line:forestcfcm-diag-bfs-begin}
        \For{\(r'=1,2,\dots,\ceil{\log_2r}\)}{
        \ForPar{\(i=1,2,\dots,2^{r'}\)}{
        \(\setof{\parent{v}|v\in\complset[\setof{s}]},\dfslist\gets\)\algname{RandomForest}\((\gr,\setof{s})\)\;\label{line:forestcfcm-diag-forest-begin}
        \ForEach{\(u\in\dfslist\)}{
            Update \(\forestfreq{u}{\setof{s}}{u}{\parent{u}}\) and \(\sum_{v\in\complset[\setof{s}]}\forestfreq{v}{\setof{s}}{u}{\parent{u}}\)\;\label{line:forestcfcm-diag-forest-end}
        }
        }
        \ForEach{\(u\in\bfslist\)}{
            \ifthenelse{\isundefined{\IEEEauthorblockN}}
            {
                \(x_u\gets\meanvolt{u}{u}{\setof{s}}-\frac{2}{n}\meanvolts{\vecone}{u}{\setof{s}}\)\tcp*{Estimator of \(\matentry{\lap^{\dagger}}{u}{u}\)}
            }
            {
                \tcp{Estimator of \(\matentry{\lap^{\dagger}}{u}{u}\)}
                \(x_u\gets\meanvolt{u}{u}{\setof{s}}-\frac{2}{n}\meanvolts{\vecone}{u}{\setof{s}}\)
            }
        }
        \(x_s\gets0\)\;\label{line:forestcfcm-approx-init-end}\label{line:forestcfcm-diag-bfs-end}
        Compute additive error bound \(\epsilon'_u\) for \(x_u\) by \lemref{lem:bernstein}\;
        \lIf{\(\epsilon'_u\leq\epsilon\normspar{x_u-\epsilon'_u}\)}{break}
        }
        \(S_1\gets\setof{\argmin_{u\in V}\setof{x_u}}\)\;
        \For{\(i=1,2,\dots,k-1\)}{
            \(\setof{\approxgain{u}{S_i}|u\in\complset[S_i]}\gets\)\algname{ForestDelta}\((\gr,S_i,\epsilon)\)\;
            \(u^*\gets\argmax_{u\in\complset[S_i]}\setof{\approxgain{u}{S_i}}\)\;
            \(S_{i+1}\gets S_i\cup\setof{u^*}\)\;
        }
        \Return{\(S_k\)}
    \end{small}
\end{algorithm}
\subsubsection{\algname{ForestCFCM}}
Given an \(n\)-node graph \(\gr\), an integer \(k\ll n\), and an error parameter \(\epsilon\), \algref{alg:forestcfcm} iteratively adds nodes to the returning set \(S\) based on greedy selection. In the first iteration, the node with minimum CFCC is chosen (\linerangeref{line:forestcfcm-approx-init-begin}{line:forestcfcm-approx-init-end}). To estimate CFCC for each node, \algname{ForestCFCM} also samples rooted spanning forests (\linerangeref{line:forestcfcm-diag-forest-begin}{line:forestcfcm-diag-forest-end}). Using counters for sampled forests, \algname{ForestCFCM} calculates the unbiased estimator described in \secref{subsec:estim-pseudolap} via BFS order (\linerangeref{line:forestcfcm-diag-bfs-begin}{line:forestcfcm-diag-bfs-end}). After obtaining the first node, \algname{ForestCFCM} repeatedly invokes \algname{ForestDelta} to estimate \(\gain{u}{S}\) for \(u\in\complset\), selecting node \(u^*\) with maximum \(\approxgain{u}{S}\). \thmref{thm:perf-forestcfcm} analyzes the performance of \algname{ForestCFCM}. Note that this nearly-linear time complexity is conservative due to \lemref{lem:bernstein}.

\begin{theorem}\label{thm:perf-forestcfcm}
    For a realistic graph \(\gr\) with \(n\) nodes, the time complexity for \algref{alg:forestcfcm} is \(\tildeo{k\epsilon^{-2}n\tau^2\dmax^{2\tau+2}}\). \algref{alg:forestcfcm} returns \(S_k\) as the approximate solution of CFCM that satisfies
    \begin{align*}
        (1+\epsilon)\bigtrace{\invsublap[u^*]}-\bigtrace{\invsublap[S_k]}                        & \geq                                                    \\
        \bigspar{1-\frac{k}{k-1}\frac{1}{\rme}-\epsilon}                   \bigspar{(1+\epsilon) & \bigtrace{\invsublap[u^*]}-\bigtrace{\invsublap[S^*]}},
    \end{align*}
    with probability of \(1-\nicefrac{1}{n}\), where
    \begin{equation*}
        u^*\triangleq\argmin_{u\in V}\matentry{\lap^{\dagger}}{u}{u},\ S^*\triangleq\argmin_{S\subset V,\abs{S}=k}\bigtrace{\invsublap}.
    \end{equation*}
\end{theorem}
\begin{IEEEproof}
    The time complexity and error guarantee follow directly from the lower bound \(\matentry{\lap^{\dagger}}{u}{u} \geq \dmax^{-1}(1-\nicefrac{1}{n})^{2}\)~\cite{VaDeCe17} as well as Lemmas~\ref{lem:perf-randomforest} and~\ref{lem:perf-forestdelta}. Plugging \lemref{lem:perf-forestdelta} into the supermodularity of \(\bigtrace{\invsublap}\), we derive for any iteration \(i\):
    \begin{equation*}
        \bigtrace{\invsublap[S_i]} - \bigtrace{\invsublap[S_{i+1}]} \geq \frac{1-\epsilon}{k}\left(\bigtrace{\invsublap[S_i]} - \bigtrace{\invsublap[S^*]}\right).
    \end{equation*}
    Recursive application yields:
    \begin{align*}
             & \bigtrace{\invsublap[S_k]}-\bigtrace{\invsublap[S^*]}                                                          \\
        \leq & \bigspar{1-\frac{1-\epsilon}{k}}^{k-1}\bigspar{\bigtrace{\invsublap[S_1]}-\bigtrace{\invsublap[S^*]}}          \\
        \leq & \bigspar{\frac{k}{k-1}\frac{1}{\rme}+\epsilon}\bigspar{\bigtrace{\invsublap[S_1]}-\bigtrace{\invsublap[S^*]}}.
    \end{align*}
    Combining this with \(\bigtrace{\invsublap[S_1]} \leq (1+\epsilon)\bigtrace{\invsublap[u^*]}\) completes the proof.
\end{IEEEproof}

\section{Algorithm Based on Schur Complement}

To address the scalability and accuracy limitations of \algname{ForestCFCM} in large networks, we present an enhanced algorithm \algname{SchurCFCM}.
The key motivation lies in leveraging the Schur complement to introduce an auxiliary root set \(T\). By analyzing the Laplacian submatrix \(\invsublap[S\cup T]\), we observe two advantages. First, the entrywise monotonicity of \(\invsublap\)~\cite{LiPeShYiZh19} ensures reduced complexity of \algname{RandomForest}. Second, \(\invsublap[S\cup T]\) exhibits stronger diagonal dominance compared to \(\invsublap\), which enhances approximation accuracy.

This section proceeds as follows. We first introduce the concept of Schur complement to improve sampling efficiency by leveraging higher-degree nodes in \(T\) (\lemref{lem:schur-sublap}). Next, we establish connections between the Schur complement and rooted probabilities in spanning forests (\lemref{lem:rooted-prob}), enabling unbiased estimation of the Schur complement. We then derive spectral sparsification guarantees (\lemref{lem:spectr-sparsif}) to ensure accurate approximation with reduced complexity. Finally, we integrate these innovations into \algname{SchurCFCM}, proving its complexity and approximation guarantee (\thmref{thm:perf-schurcfcm}).

\begin{comment}
While \algname{ForestCFCM} solves CFCM with an approximation factor, both its efficiency and effectiveness can be improved. Specifically, when the root set \(S\) of a large-scale network contains only a few nodes, it becomes time-consuming for a random walker to reach an arbitrary node in \(S\), leading to inefficient forest sampling. Moreover, as discussed in \secref{subsec:estim-pseudolap}, the tiny capacity of \(S\) results in relatively weak diagonal dominance, which contributes to the inaccuracy of approximation algorithms. Denoting a node subset of \(\complset\) as \(T\), \lemref{lem:sublap-pseudolap} motivates us to represent \(\invsublap\) in terms of \(\invsublap[(S\cup T)]\) to further enhance the efficiency and effectiveness of solving CFCM. For this purpose, we introduce the notion of Schur complement to design an improved approximation algorithm.
\end{comment}

\subsection{Schur Complement and Its Properties}

\begin{definition}[Schur complement]\label{def:schur-compl}
    For a square matrix \(\matm\), we can adjust its index order and rewrite \(\matm\) in block form as \(\matm=\begin{pmatrix}\matentry{\matm}{U}{U} & \matentry{\matm}{U}{T} \\\matentry{\matm}{T}{U} & \matentry{\matm}{T}{T}\end{pmatrix}\). The Schur complement of \(\matm\) onto index subset \(T\) is then defined as \(\schcompl{\matm}{T}\triangleq\matentry{\matm}{T}{T}-\matentry{\matm}{T}{U}\matentry{\matm}{U}{U}^{-1}\matentry{\matm}{U}{T}\).
\end{definition}

For an undirected graph \(\gr=(V,E)\) with non-empty node subsets \(S\subset V\) and \(T\subset\complset\), we denote \(\complset[(S\cup T)]\) as \(U\). Then \(\invsublap\) can be represented as~\cite{Zh06}:
\begin{equation}\label{eq:inv-schur-compl}
    \small
    \invsublap=
    \begin{pmatrix}
        \invmatentry{\lap}{U}{U}+\matf\bigparinv{\schcompl{\sublap}{T}}\matf^\top & \matf\bigparinv{\schcompl{\sublap}{T}} \\
        \bigparinv{\schcompl{\sublap}{T}}\matf^\top                               & \bigparinv{\schcompl{\sublap}{T}}
    \end{pmatrix},
\end{equation}
where \(\matf\) denotes the matrix \(-\invmatentry{\lap}{U}{U}\matentry{\lap}{U}{T}\).

Equation~\eqref{eq:inv-schur-compl} shows that the computation of \(\invsublap\) can be transformed into calculations involving \(\invmatentry{\lap}{U}{U}\), \(\matf\), and \(\bigparinv{\schcompl{\sublap}{T}}\). While \(\invmatentry{\lap}{U}{U}\) can be estimated by sampling spanning forests with root set \(S\cup T\), we will demonstrate that \(\matf\) and \(\schcompl{\sublap}{T}\) can also be represented by quantities related to random spanning forests.
First, we derive the connection between \(\matf\) and rooted probability.

\begin{lemma}\label{lem:rooted-prob}
    For random spanning forests of graph \(\gr\) with root set \(S\cup T\), let \(\rootprob{u}{t}\) denote the probability that \(u\in U\) belongs to the tree whose root is \(t\in T\). Then we have
    \begin{equation}\label{eq:root-prob}
        \matentry{\matf}{u}{t}=-\vece_u^\top\invmatentry{\lap}{U}{U}\matentry{\lap}{U}{T}\vece_t=\rootprob{u}{t}.
    \end{equation}
\end{lemma}
\begin{IEEEproof}
    Following the interpretation of absorbing random walk~\cite{KeSn76}, \(\matentry{\matf}{u}{t}\) equals the probability that a walk starting at \(u\) is absorbed at \(t \in T\). In the initial round of loop-erased random walk from node \(u\), the walk behaves identically to an absorbing random walk with target set \(T\). By Wilson's algorithm~\cite{Wi96}, the sampling distribution is invariant to the order of source nodes, thus \(\matentry{\matf}{u}{t} = \rootprob{u}{t}\) holds for all \(u \in U\).
\end{IEEEproof}

Motivated by \lemref{lem:rooted-prob}, we define an unbiased estimator of \(\matentry{\matf}{u}{t}\). Let \(\tildeforestmark(S\cup T)\) be the number of sampled spanning forests with root set \(S\cup T\), and \(\rootnum{u}{t}\) denote the number of these forests where node \(u\) is rooted at \(t\in T\). Then \(\rootnum{u}{t}/\tildeforestmark(S\cup T)\) serves as an unbiased estimator of \(\matentry{\matf}{u}{t}\).

Next, we address the approximation of \(\bigparinv{\schcompl{\sublap}{T}}\in\rea^{\abs{T}\times\abs{T}}\). Given that \(\abs{T}\ll\abs{V}\) is still relatively small, we can efficiently compute the inverse matrix if \(\schcompl{\sublap}{T}\) is obtained. To achieve this, we first provide a relationship between \(\schcompl{\lap}{T}\) and the Laplacian matrix of another graph, then extend this relationship to the case of \(\schcompl{\sublap}{T}\).

For a graph \(\gr=(V,E)\) with non-empty node subset \(T\subset V\), the Schur complement \(\schcompl{\lap}{T}\) is also the Laplacian matrix of a weighted graph denoted as \(\schcompl{\gr}{T}\), whose node set is precisely \(T\)~\cite{De22}. Regarding the Laplacian submatrix \(\sublap\), we establish a connection between \(\schcompl{\sublap}{T}\) and \(\schcompl{\gr}{(S\cup T)}\).
\begin{lemma}\label{lem:schur-sublap}
    For a graph \(\gr=(V,E)\) with non-empty node subsets \(S\subset V\) and \(T\subset\complset\), the Schur complement of \(\sublap\) onto \(T\) is equivalent to the Laplacian submatrix of \(\schcompl{\gr}{(S\cup T)}\):
    \begin{equation*}
        \schcompl{\sublap}{T}=\bigspar{\schcompl{\lap}{(S\cup T)}}_{\scriptscriptstyle -S}.
    \end{equation*}
\end{lemma}
\begin{IEEEproof}
    From~\cite{De22}, \(\schcompl{\lap}{(S\cup T)}\) is the Laplacian matrix of weighted graph \(\schcompl{\gr}{(S\cup T)}\), which can be rewritten as
    \begin{equation}\label{eq:schcompl-lap}
        \schcompl{\lap}{(S\cup T)}=\matentry{\lap}{(S\cup T)}{(S\cup T)}-\matentry{\lap}{(S\cup T)}{U}\matentry{\lap^{-1}}{U}{U}\matentry{\lap}{U}{(S\cup T)}.
    \end{equation}
    Additionally, \(\schcompl{\sublap}{T}\) can also be represented as
    \begin{equation}\label{eq:schcompl-sublap}
        \schcompl{\sublap}{T}=\matentry{\lap}{T}{T}-\matentry{\lap}{T}{U}\matentry{\lap^{-1}}{U}{U}\matentry{\lap}{U}{T}.
    \end{equation}
    Combining~\eqref{eq:schcompl-lap} with~\eqref{eq:schcompl-sublap}, we observe that \(\schcompl{\sublap}{T}\) can be obtained from \(\schcompl{\lap}{(S\cup T)}\) by removing elements with row indices or column indices in set \(S\), completing the proof.
\end{IEEEproof}

According to \defref{def:schur-compl}, \(\schcompl{\sublap}{T}\) can also be represented as \(\matentry{\lap}{T}{T}+\matentry{\lap}{T}{U}\matf\). Combining this expression with \lemref{lem:rooted-prob}, the entry \((i,j)\) of \(\schcompl{\sublap}{T}\) can be rewritten as
\begin{equation}\label{eq:element-schur-compl}
    \small
    \vece_i^\top\schcompl{\sublap}{T}\vece_j=
    \begin{cases}
        \matentry{\lap}{i}{i}-\sum_{\edge{u}{i}\in E}\rootprob{u}{i} & i=j     \\
        \matentry{\lap}{i}{j}-\sum_{\edge{u}{i}\in E}\rootprob{u}{j} & i\neq j
    \end{cases}.
\end{equation}
By integrating \lemref{lem:rooted-prob} with~\eqref{eq:inv-schur-compl} and~\eqref{eq:element-schur-compl}, we finally demonstrate that \(\invsublap\) can be estimated by sampling spanning forests with root set \(S\cup T\), incorporating the additional approximation of rooted probabilities. \lemref{lem:perf-randomforest} suggests that replacing root set \(S\) with \(S\cup T\) enhances the efficiency of forest sampling.

\subsection{Algorithm Design and Analyses}\label{subsec:alg-schurcfcm}

Based on the preceding analyses, we propose a more efficient algorithm \algname{SchurCFCM} for solving CFCM, which is depicted in \algref{alg:schurcfcm}. In each subsequent iteration, \algname{SchurCFCM} invokes \algname{SchurDelta}, as shown in \algref{alg:schurdelta}. Both of these algorithms utilize \algname{RandomForest} to sample random spanning forests, which is illustrated in \algref{alg:randomforest}.

\begin{algorithm}
    \caption{\algname{SchurDelta}\((\gr,S,T,\epsilon)\)}
    \label{alg:schurdelta}
    \begin{small}
        \Input{
            \(\gr=(V,E)\): an undirected graph;
            \(S\subset V\): the node subset;
            \(T\subset \complset\): the additional root set;
            \(\epsilon\): an error parameter
        }
        \Output{
            \(\approxgain{u}{S}\): the margin for node \(u\in\complset\)
        }
        \(U\gets\complset[(S\cup T)]\)\;
        \(w\gets\ceil{24(\nicefrac{\epsilon}{7})^{-2}\log n}\), \(r\gets{8(\nicefrac{\epsilon}{15})^{-2}\tau^2\dmax^{2\tau+2}(S\cup T)\log(2n)}\)\;
        \(\forestfreq{u}{S\cup T}{a}{b}\gets 0\) for \(u\in U\) and \(\edge{a}{b}\in E\)\;\label{line:schurdelta-forest-begin}
        Construct \(\matw\in\rea^{w\times \abs{U}}\) and \(\matq\in\rea^{w\times \abs{T}}\) by \lemref{lem:jl}\;
        \(\bfslist\gets\) the list of nodes visited in BFS order from \(S\cup T\)\;\label{line:schurdelta-bfs-begin}
        \For{\(r'=1,2,\dots,\ceil{\log_2r}\)}{
        \ForPar{\(i=1,2,\dots,2^{r'}\)}{
        \(\setof{\parent{u}|u\in U},\dfslist\gets\)\algname{RandomForest}\((\gr,S\cup T)\)\;
        \ForEach{\(u\in\dfslist\)}{
            Update \(\sum_{v\in U}\matentry{\matw}{j}{v}\forestfreq{v}{S\cup T}{u}{\parent{u}}\) for \(j=1,2,\dots,w\)\;\label{line:schurdelta-count-begin}
            \ifthenelse{\isundefined{\IEEEauthorblockN}}
            {
                Update \(\forestfreq{u}{S\cup T}{u}{\parent{u}}\)
            }
            {
                Update \(\forestfreq{u}{S\cup T}{u}{\parent{u}}\)
            }\label{line:schurdelta-forest-end}\label{line:schurdelta-count-end}
        }
        }
        \(\matentry{\tildematf}{u}{t}\gets\rootnum{u}{t}/\tildeforestmark(S\cup T)\) for \(u\in U,v\in T\)\;
        construct \(\approxschcompl{\sublap}{T}\) by~\eqref{eq:element-schur-compl}\;\label{line:schurdelta-schur-compl}
        \(\maty\in\rea^{w\times\abs{\complset}}\gets\) arbitrary value\;
        \ForEach{\(u\in\bfslist\)}{
            \(\matentry{\maty}{j}{u}\gets\meanvolts{\vece_j^\top\matw}{u}{S\cup T}+\vece_j^\top\normspar{\matw\tildematf+\matq}\bigparinv{\approxschcompl{\sublap}{T}}\tildematf\vece_u\) for \(j=1,2,\dots,w\)\;
            \(z_u\gets\meanvolt{u}{u}{S\cup T}+\vece_u^\top\tildematf\bigparinv{\approxschcompl{\sublap}{T}}\tildematf\vece_u\)\;\label{line:schurdelta-bfs-end}
        }
        \(\maty\vece_t\gets\normspar{\matw\tildematf+\matq}\bigparinv{\approxschcompl{\sublap}{T}}\vece_t\) for \(t\in T\)\;
        \(z_t\gets\vece_t^\top\bigparinv{\approxschcompl{\sublap}{T}}\vece_t\) for \(t\in T\)\;
        \(\approxgain{u}{S}\gets z_u^{-1}\Abs{\maty\vece_u}^2\) for \(u\in\complset\)\;
        Compute error bound \(\epsilon'_u\) for \(\approxgain{u}{S}\) by \lemref{lem:bernstein}\;
        \lIf{\({\epsilon'_u}\leq\epsilon\normspar{\approxgain{u}{S}-\epsilon'_u}\) for \(u\in V\setminus S\)}{break}
        }
        \Return{\(\setof{\approxgain{u}{S}=z_u^{-1}\Abs{\maty\vece_u}^2\big\vert u\in\complset}\)}
    \end{small}
\end{algorithm}
\subsubsection{\algname{SchurDelta}}
While \algname{SchurDelta} shares similarities with \algname{ForestDelta} in estimating \(\gain{u}{S}\), it additionally accepts the additional root set \(T\). Therefore, the root set of sampled spanning forest becomes \(S\cup T\). Given that \(\invsublap\) has been rewritten as~\eqref{eq:inv-schur-compl}, we need to approximate the matrix of rooted probability \(\tildematf\), which can be efficiently maintained during the sampling process. Furthermore, we estimate the Schur complement \(\schcompl{\sublap}{T}\) based on \(\tildematf\) (\lineref{line:schurdelta-schur-compl}). As \(\invsublap\) has been represented in block form, we finally calculate the estimated value of \(\gain{u}{S}\) for \(u\in U\) and \(u\in T\) respectively. In order to reduce the sample size \(r\) without affecting the theoretical guarantee, we still split it into \(\ceil{\log_2 r}\) batches, utilizing \lemref{lem:bernstein} for possible early termination. To analyze the theoretical accuracy of \algname{SchurDelta}, we first introduce the following lemma~\cite{SpSr08}.

\begin{lemma}\label{lem:spectr-sparsif}
    Let \(\randgr{\tilgr}{i}\) denote the \(\myord{i}\) random multi-subgraph of \(\gr\) with \(n\) nodes, where \(i=1,2,\dots,r\). Each generated edge \(\edge{u}{v}\) in \(\randgr{\tilgr}{i}\) has weight \(\nicefrac{1}{r}\). Let \(\tilgr\) be the weighted graph obtained by summing the graphs \(\randgr{\tilgr}{i}\). For edge \(\edge{a}{b}\) in \(\tilgr\), its weight is equivalent to the sum of weight of every edge \(\edge{a}{b}\) in \(\randgr{\tilgr}{i}\). If \(\tilgr\) is an unbiased estimator of \(\gr\) and the weight of any edge \(\edge{a}{b}\in\randgr{\tilgr}{i}\) is less than \(\frac{\epsilon^2}{\resist{a}{b}\log n}\), then \(\tilgr\) is an \(\epsilon\)-spectral sparsifier of \(\gr\) with high probability. In other words, \(\vecx^\top\tillap\vecx\approx_{\epsilon}\vecx^\top\lap\vecx\) holds for any real vector \(\vecx\), where \(\tillap\) and \(\lap\) are Laplacian matrices of \(\tilgr\) and \(\gr\), respectively.
\end{lemma}

Next, the relative error guarantee of \algname{SchurDelta} is given in \lemref{lem:error-schurdelta}.

\begin{lemma}\label{lem:error-schurdelta}
    For an undirected graph \(\gr\) and the error parameter \(\epsilon\), if the sample size \(r\) satisfies
    \begin{equation}\label{eq:error-schurdelta}
        r\geq8(\nicefrac{\epsilon}{15})^{-2}\tau^2\dmax^{2\tau+2}(S\cup T)\log(2n),
    \end{equation}
    \algref{alg:schurdelta} returns \(\approxgain{u}{S}\approx_{\epsilon}\gain{u}{S}\) for \(u\in V\setminus S\) with probability of \(1-\nicefrac{1}{n}\).
\end{lemma}
\begin{IEEEproof}
    We establish the relative error guarantee through three key steps.

    \textit{Step 1: Spectral Sparsification of Schur Complement.}
    From \lemref{lem:spectr-sparsif}, an \(\nicefrac{\epsilon}{15}\)-spectral sparsifier \(\approxschcompl{\sublap}{T}\) of \(\schcompl{\sublap}{T}\) can be constructed with high probability when the sample size satisfies~\eqref{eq:error-schurdelta}. This follows because:
    (i) Edge weights in random subgraphs are bounded by \(\frac{\dmax(S\cup T)}{r}\) via~\eqref{eq:element-schur-compl},
    (ii) Maximum resistance distance \(\resist{a}{b} \leq \tau\) ensures that edge weights meet the condition \(\frac{\normspar{\nicefrac{\epsilon}{15}}^2}{\resist{a}{b}\log n} \geq \frac{\dmax(S\cup T)}{r}\).

    \textit{Step 2: Denominator Estimation.}
    Equation~\eqref{eq:inv-schur-compl} decomposes the denominator \(\matentry{\bigspar{\invsublap}}{u}{u}\) into three terms:.
    \begin{align*}
        \matentry{\bigspar{\invsublap}}{u}{u}
        = & \matentry{\bigspar{\invmatentry{\lap}{U}{U}}}{u}{u} + \matentry{\bigspar{\matf\bigparinv{\schcompl{\sublap}{T}}\matf^\top}}{u}{u} \\
          & + \matentry{\bigspar{\bigparinv{\schcompl{\sublap}{T}}}}{u}{u}.
    \end{align*}
    For the first term, the sample size satisfying~\eqref{eq:error-schurdelta} yields \(\meanvolt{u}{u}{S\cup T} \approx_{\epsilon/3} \matentry{\bigspar{\invmatentry{\lap}{U}{U}}}{u}{u}\) by \lemref{lem:perf-forestdelta}. For the second term, let \(X=\matentry{\bigspar{\bigparinv{\schcompl{\sublap}{T}}}}{t}{t}\) if \(u\) roots at \(t \in T\). Then \(\mathbb{E}[X] = \matentry{\bigspar{\matf\bigparinv{\schcompl{\sublap}{T}}\matf^\top}}{u}{u}\) and \(X \leq (1+\nicefrac{\epsilon}{3})\tau\) via spectral sparsification. Applying \lemref{lem:hoeffding} with~\eqref{eq:error-schurdelta} yields
    \begin{equation*}
        \matentry{\bigspar{\tildematf\bigparinv{\approxschcompl{\sublap}{T}}\tildematf^\top}}{u}{u}\approx_{\nicefrac{\epsilon}{3}}\matentry{\bigspar{\matf\bigparinv{\schcompl{\sublap}{T}}\matf^\top}}{u}{u}.
    \end{equation*}
    The third term directly follows from \lemref{lem:spectr-sparsif}.

    \textit{Step 3: Numerator Estimation.}
    The numerator \(\|\invsublap\vece_u\|^2\) is estimated via \lemref{lem:jl}. By~\eqref{eq:inv-schur-compl}, this estimation boils down to estimating non-diagonal elements of \(\invmatentry{\lap}{U}{U}\), \(\bigparinv{\schcompl{\sublap}{T}}\), \(\matf\bigparinv{\schcompl{\sublap}{T}}\) and \(\matf\bigparinv{\schcompl{\sublap}{T}}\matf^\top\), which are similar with aforementioned diagonal ones, except for two differences:
    (i) Upper bound \((1+\nicefrac{2\epsilon}{15})\tau\) of \(\matentry{\approxschcompl{\sublap}{T}}{t_1}{t_2}\) derived from spectral sparsification,
    (ii) Lower bound \(\dmax^{-\tau-1}(S\cup T)\) for \(\matentry{\bigspar{\invmatentry{\lap}{U}{U}}}{u_1}{u_2}\) derived from Neumann series expansion.

    \textit{Error Composition.}
    Combining these estimations with \(w\geq24(\nicefrac{\epsilon}{7})^{-2}\log n\) from \lemref{lem:jl}, we have
    \begin{align*}
        \frac{|\approxgain{u}{S} - \gain{u}{S}|}{\gain{u}{S}}
         & \leq \frac{(1+\nicefrac{\epsilon}{15})^2(1+\nicefrac{\epsilon}{7})}{1 - \nicefrac{\epsilon}{3}} - 1 \\
         & \leq \epsilon \quad \text{for } 0 < \epsilon < 1.
    \end{align*}
    Thus, \(\approxgain{u}{S} \approx_{\epsilon} \gain{u}{S}\) holds with probability \(1 - \nicefrac{1}{n}\).
\end{IEEEproof}

Finally, the performance of \algname{SchurDelta} is analyzed:

\begin{lemma}\label{lem:perf-schurdelta}
    For a graph \(\gr\), the time complexity for \algref{alg:schurdelta} is \(\bigo{\mathcal{T}}\), where
    \begin{align*}
        \mathcal{T}=\
         & \epsilon^{-2}\tau^2\dmax^{2\tau+2}(S\cup T)\log n\bigtrace{\bigparinv{\mati-\matp_{-(S\cup T)}}} \\
         & +\epsilon^{-2}n\log n.
    \end{align*}
    Specifically, if \(\gr\) is a real-world graph, this upper bound can be expressed as \(\tildeo{\epsilon^{-2}n\tau^2}\).
\end{lemma}
\begin{IEEEproof}
    Similar to the proof of \lemref{lem:perf-forestdelta}, we omit analyzing the time complexity of \algname{SchurDelta} being \(\normo{\mathcal{T}}\). We next prove that for real-world graphs, the time complexity of \algname{SchurDelta} scales nearly linearly in the number of nodes.

    According to \tabref{tab:notation}, \(\dmax(S\cup T)\) denotes the maximum degree after removing nodes in \(S\cup T\). Due to the scale-free property of real-world graphs~\cite{AlBa02}, removing hub nodes from \(T\) reduces \(\dmax(S\cup T)\) under the average degree, which is typically a small constant for realistic graphs. The empirical upper bound of \(\dmax(S\cup T)\) is presented in attribute \(\abs{T*}\) of \tabref{tab:time}, validating the above analysis. Meanwhile, the diameter \(\tau\) of scale-free graphs is \(\normo{\log n}\) or even \(\normo{\log\log n}\)~\cite{LeKlFa05}. Therefore, \(\tau^2\dmax^{2\tau+2}(S\cup T)\) is very small, even less than a poly-log factor for real graphs, preserving the nearly-linear time complexity of \algref{alg:schurdelta}.
\end{IEEEproof}

Despite sharing the approach of sampling spanning forests with \algname{ForestDelta}, \algname{SchurDelta} exhibits superior theoretical performance, which stems from the additional root set \(T\). As \(\matentry{\bigparinv{\mati-\matp_{-(S\cup T)}}}{i}{i}\) can be rewritten as \(d_i\matentry{\spar{\invsublap[S\cup T]}}{i}{i}\), The entrywise monotonicity of \(\invsublap\)~\cite{LiPeShYiZh19} greatly reduces the complexity of invoking \algref{alg:randomforest} when \(T\) contains a few nodes. Meanwhile, \(\invsublap[S\cup T]\) has better diagonal dominance than \(\invsublap\), leading to better accuracy of \algname{ForestDelta}.

\begin{algorithm}
    \caption{\algname{SchurCFCM}\((\gr,k,\epsilon)\)}
    \label{alg:schurcfcm}
    \begin{small}
        \Input{
            \(\gr=(V,E)\): an undirected graph;
            \(k\ll\abs{V}\): the capacity of node set;
            \(c\ll\abs{V}\): the capacity of additional root set;
            \(\epsilon\): an error parameter
        }
        \Output{
            \(S_k\): A subset of \(V\) with \(\abs{S_k}=k\)
        }
        Select \(c\) nodes with maximum degree to form \(T\)\;
        \(s\gets\argmax_{u\in V}d_u\), \(r\gets{18\epsilon^{-2}\tau^2d_s^2(1-\nicefrac{1}{n})^{-4}\log(2n)}\)\;
        \(\forestfreq{u}{\setof{s}}{a}{b}\gets 0\) for \(u\in\complset[\setof{s}]\) and \(\edge{a}{b}\in E\)\;\label{line:schurcfcm-approx-init-begin}
        \(\bfslist\gets\) the list of nodes visited in BFS order from \(s\)\;\label{line:schurcfcm-diag-bfs-begin}
        \For{\(r'=1,2,\dots,\ceil{\log_2r}\)}{
        \ForPar{\(i=1,2,\dots,2^{r'}\)}{
        \(\setof{\parent{v}|v\in\complset[\setof{s}]},\dfslist\gets\)\algname{RandomForest}\((\gr,\setof{s})\)\;\label{line:schurcfcm-diag-forest-begin}
        \ForEach{\(u\in\dfslist\)}{
            Update \(\forestfreq{u}{\setof{s}}{u}{\parent{u}}\) and \(\sum_{v\in\complset[\setof{s}]}\forestfreq{v}{\setof{s}}{u}{\parent{u}}\)\;\label{line:schurcfcm-diag-forest-end}
        }
        }
        \ForEach{\(u\in\bfslist\)}{
            \tcp{Estimator of \(\matentry{\lap^{\dagger}}{u}{u}\)}\(x_u\gets\meanvolt{u}{u}{\setof{s}}-\frac{2}{\abs{V}}\meanvolts{\vecone}{u}{\setof{s}}\)
        }
        \(x_s\gets0\)\;\label{line:schurcfcm-approx-init-end}\label{line:schurcfcm-diag-bfs-end}
        Compute additive error bound \(\epsilon'_u\) for \(x_u\) by \lemref{lem:bernstein}\;
        \lIf{\(\epsilon'_u\leq\epsilon\normspar{x_u-\epsilon'_u}\)}{break}
        }
        \(S_1\gets\setof{\argmin_{u\in V}\setof{x_u}}\)\;
        \For{\(i=1,2,\dots,k-1\)}{
            \(\setof{\approxgain{u}{S_i}|u\in\complset[S_i]}\gets\)\algname{SchurDelta}\((\gr,S_i,T\setminus S_i,\epsilon)\)\;
            \(u^*\gets\argmax_{u\in\complset[S_i]}\setof{\approxgain{u}{S_i}}\)\;
            \(S_{i+1}\gets S_i\cup\setof{u^*}\)\;
        }
        \Return{\(S_k\)}
    \end{small}
\end{algorithm}
\subsubsection{\algname{SchurCFCM}}
Due to the leverage of \lemref{lem:jl}, the performance bottleneck stems from \algname{SchurDelta}. We thus do not introduce Schur complement in \algref{alg:schurcfcm} for ease of implementation. However, it remains necessary to determine the nodes of additional root set \(T\). According to the proof of \lemref{lem:perf-randomforest}, the upper bound of time complexity for sampling a spanning forest depends on the mean hitting time to \(S\cup T\). Enhanced reachability of \(T\) leads to improved efficiency of \algname{SchurCFCM}. Therefore, we repeatedly select the node with maximum degree in the remaining graph, which is both reasonable and efficient for implementation. The performance of \algname{SchurCFCM} is characterized in \thmref{thm:perf-schurcfcm}. Similar to the analysis of \algname{ForestCFCM}, this upper bound of time complexity is also conservative due to the application of \lemref{lem:bernstein}.

\begin{theorem}\label{thm:perf-schurcfcm}
    For a real-world \(n\)-node graph \(\gr\), the upper bound of the time complexity for \algref{alg:schurcfcm} is \(\tildeo{k\epsilon^{-2}n\tau^2\dmax^{2\tau+2}}\). \algref{alg:schurcfcm} returns \(S_k\) as the approximate solution of CFCM that satisfies
    \begin{align*}
        (1+\epsilon)\bigtrace{\invsublap[u^*]}-\bigtrace{\invsublap[S_k]}                        & \geq                                                   \\
        \biggspar{1-\frac{k}{k-1}\frac{1}{\rme}-\epsilon}                  \bigspar{(1+\epsilon) & \bigtrace{\invsublap[u^*]}-\bigtrace{\invsublap[S^*]}}
    \end{align*}
    with probability of \(1-\nicefrac{1}{n}\), where
    \begin{equation*}
        u^*\triangleq\argmin_{u\in V}\matentry{\lap^{\dagger}}{u}{u},\ S^*\triangleq\argmin_{S\subset V,\abs{S}=k}\bigtrace{\invsublap}.
    \end{equation*}
\end{theorem}
\begin{IEEEproof}
    The time complexity of \algname{ForestCFCM} and the relative error guarantee for \(\matentry{\lap^{\dagger}}{u}{u}\) follow directly from combining Lemmas~\ref{lem:perf-randomforest} and~\ref{lem:perf-forestdelta}. To derive the approximation factor of solving CFCM, we observe that \algname{SchurCFCM} and \algname{ForestCFCM} differ only in their iterative subroutine: \algname{SchurDelta} for the former and \algname{ForestDelta} for the latter. As both subroutines provide an \(\epsilon\)-approximation of \(\gain{u}{S}\), the proof of approximation factor for \algname{SchurCFCM} mirrors that of \algname{ForestCFCM}.
\end{IEEEproof}

\section{Numerical Experiments}\label{sec:num-exp}

\subsection{Experimental Settings}\label{subsec:exp-setting}

\textbf{Datasets.}
Our experiments utilize real-world graph data from KONECT~\cite{Ku13}, SNAP~\cite{LeKr14}, and Network Repository~\cite{RoAh15}. To ensure fair comparison with previous works, we also conduct experiments on several networks examined in~\cite{LiPeShYiZh19}. For networks that are not originally connected, we perform our experiments on their largest connected components (LCCs). \tabref{tab:time} presents relevant information about the LCCs of the studied real-world networks, listed in ascending order by node number. The smallest network contains 1039 nodes, while the largest comprises 6 million nodes.

\begin{table*}
    \centering
    % \tabcolsep=3pt
    % \fontsize{7.0}{7.3}\selectfont
    \def\exa{\algname{Exact}}
    \def\app{\algname{Approx}}
    \def\for{\algname{ForestCFCM}}
    \def\sch{\algname{SchurCFCM}}
    \caption{The running time (seconds, \(s\)) of {\exa}, \algname{ApproxGreedy} (\app), \algname{ForestCFCM} and \algname{SchurCFCM} with various \(\epsilon\) on real-world graphs.}
    \label{tab:time}
    \begin{tabular}{@{}crrccrrcccccc@{}}
        \toprule
        \multirow{3}{*}{Network}
                          &
        \multirow{3}{*}{Node}
                          &
        \multirow{3}{*}{Edge}
                          &
        \multirow{3}{*}{\(\tau\)}
                          &
        \multirow{3}{*}{\(\abs{T^*}\)}
                          &
        \multicolumn{8}{c}{Running time (seconds)}                                                                                \\
        \cmidrule(l){6-13}
                          &           &             &    &      &
        \multirow{2}{*}{\exa}
                          &
        \multirow{2}{*}{\app}
                          &
        \multicolumn{3}{c}{\for}
                          &
        \multicolumn{3}{c}{\sch}                                                                                                  \\
        \cmidrule(l){8-13}
                          &           &             &    &      &        &        &
        \(\epsilon=0.3\)  &
        \(\epsilon=0.2\)  &
        \(\epsilon=0.15\) &
        \(\epsilon=0.3\)  &
        \(\epsilon=0.2\)  &
        \(\epsilon=0.15\)                                                                                                         \\
        \midrule
        Euroroads         & 1,039     & 1,305       & 62 & 7    & 4.824  & 8.491  & 0.328 & 0.497 & 0.825 & 0.283 & 0.451 & 0.709 \\
        Hamsterster       & 2,000     & 16,097      & 10 & 58   & 33.70  & 34.43  & 0.747 & 1.273 & 1.993 & 0.532 & 0.992 & 1.659 \\
        Facebook          & 4,039     & 88,234      & 8  & 127  & 274.6  & 196.2  & 2.446 & 4.321 & 6.901 & 1.695 & 3.448 & 5.608 \\
        GR-QC             & 4,158     & 13,428      & 17 & 34   & 298.8  & 60.41  & 2.876 & 5.450 & 9.008 & 2.404 & 4.867 & 8.246 \\
        web-EPA           & 4,253     & 8,897       & 10 & 43   & 319.1  & 32.91  & 2.631 & 5.024 & 8.359 & 2.216 & 4.513 & 7.699 \\
        Routeviews        & 6,474     & 13,895      & 9  & 45   & 1130   & 39.88  & 4.440 & 8.499 & 14.21 & 3.938 & 8.029 & 13.65 \\
        soc-PagesGov      & 7,057     & 89,429      & 10 & 113  & 1455   & 253.4  & 6.371 & 11.88 & 19.94 & 5.444 & 10.33 & 17.42 \\
        HEP-Th            & 8,638     & 24,827      & 18 & 37   & 2676   & 157.4  & 8.125 & 15.76 & 25.50 & 6.679 & 13.39 & 22.76 \\
        Astro-Ph          & 17,903    & 197,031     & 14 & 138  & 24456  & 1118   & 22.10 & 44.24 & 74.35 & 18.73 & 35.69 & 59.81 \\
        CAIDA             & 26,475    & 53,381      & 17 & 86   & 81549  & 392.4  & 27.41 & 53.98 & 91.53 & 26.37 & 50.58 & 84.81 \\
        EmailEnron        & 33,696    & 180,811     & 13 & 177  & 161354 & 1247   & 48.89 & 92.60 & 147.6 & 39.50 & 79.86 & 130.3 \\
        Brightkite        & 56,739    & 212,945     & 18 & 146  & --     & 1694   & 57.17 & 98.22 & 179.0 & 56.97 & 97.52 & 154.3 \\
        buzznet           & 101,163   & 2,763,066   & 4  & 664  & --     & 10121  & 80.79 & 126.8 & 196.0 & 73.59 & 126.7 & 176.2 \\
        Livemocha         & 104,103   & 2,193,083   & 6  & 631  & --     & 12364  & 83.24 & 149.8 & 232.1 & 81.86 & 128.7 & 218.4 \\
        WordNet           & 145,145   & 656,230     & 16 & 205  & --     & 6153   & 142.6 & 243.7 & 383.1 & 112.4 & 203.0 & 333.6 \\
        Gowalla           & 196,591   & 950,327     & 16 & 258  & --     & 8900   & 180.4 & 314.8 & 497.6 & 149.0 & 271.8 & 445.3 \\
        com-DBLP          & 317,080   & 1,049,866   & 23 & 131  & --     & 13022  & 170.1 & 290.7 & 444.1 & 130.6 & 241.5 & 396.9 \\
        Amazon            & 334,863   & 925,872     & 47 & 96   & --     & 19252  & 200.0 & 356.7 & 562.9 & 172.4 & 311.2 & 519.3 \\
        Actor             & 374,511   & 15,014,839  & 13 & 1174 & --     & 100333 & 270.9 & 480.7 & 764.5 & 215.5 & 403.1 & 656.3 \\
        Dogster           & 426,485   & 8,543,321   & 11 & 1174 & --     & 43005  & 221.3 & 312.9 & 445.2 & 150.2 & 227.3 & 372.8 \\
        FourSquare        & 639,014   & 3,214,986   & 4  & 201  & --     & --     & 318.9 & 414.6 & 585.4 & 266.5 & 403.4 & 558.6 \\
        Skitter           & 1,694,616 & 11,094,209  & 31 & 965  & --     & --     & 585.4 & 774.0 & 1064  & 372.0 & 487.5 & 750.2 \\
        Flixster          & 2,523,386 & 7,918,801   & 7  & 945  & --     & --     & 642.0 & 835.2 & 1080  & 292.7 & 504.0 & 755.2 \\
        Orkut             & 2,997,166 & 106,349,209 & 9  & 1462 & --     & --     & 1104  & 1576  & 2223  & 752.3 & 1125  & 1707  \\
        Youtube           & 3,216,075 & 9,369,874   & 31 & 892  & --     & --     & 1009  & 1307  & 1796  & 618.4 & 903.4 & 1376  \\
        soc-LiveJournal   & 5,189,808 & 48,687,945  & 23 & 951  & --     & --     & 2017  & 2712  & 3693  & 1152  & 1864  & 2758  \\
        sc-rel9           & 5,921,786 & 23,667,162  & 7  & 125  & --     & --     & 1474  & 2058  & 2920  & 969.9 & 1517  & 2302  \\
        \bottomrule
    \end{tabular}
\end{table*}

\textbf{Environment.}
All experiments are conducted on a Linux server equipped with a 72-core 2.1GHz CPU and 256GB of RAM. We implement \algname{ApproxGreedy} in Julia, maintaining consistency with~\cite{LiPeShYiZh19}, which relies on a Julia-based Laplacian solver~\cite{KySa16}. Our proposed algorithms and other baseline methods are implemented in C++. Given that all algorithms are pleasingly parallelizable, we execute each experiment using 72 threads.

\textbf{Baselines and Parameters.}
To demonstrate the superiority of our algorithms, we first implement the state-of-the-art method \algname{ApproxGreedy}~\cite{LiPeShYiZh19} as a baseline. Since this method also utilizes \lemref{lem:jl}, we set the error parameter \(\epsilon\) to be \(0.2\), matching \algname{ForestCFCM} and \algname{SchurCFCM}. We also implement the greedy algorithm \algname{Exact}, which calculates \(\lap^{\dagger}\) and \(\gain{u}{S}\) through matrix inversion. To ensure that our greedy algorithms outperform other heuristic strategies in solving CFCM effectively, we implement two additional heuristic methods: \algname{Degree} and \algname{Top-CFCC}. \algname{Degree} selects \(k\) nodes with the largest degrees, while \algname{Top-CFCC} selects top-\(k\) nodes with the largest CFCC of single nodes.

While the nodes of additional root set \(T\) in \algname{SchurCFCM} are selected according to maximum degree, we also need to determine the size of additional root set \(T\). When \(\abs{T}\) becomes larger, the time of directly inverting the Schur complement grows cubically. When \(\abs{T}\) becomes smaller, the upper bound of \algname{SchurDelta} increases due to the growing \(\dmax(T)\). Therefore, we attempt to reach a balance between these two factors, setting the size as \(\abs{T^*}=\argmin_{\abs{T}}\setof{\bigabs{\abs{T}-\dmax(T)}}\). \tabref{tab:time} presents the value of \(\abs{T^*}\) for each tested graph, which is very small due to the scale-free property of real-world graphs.

\subsection{Results on Real-world Networks}

\subsubsection{Efficiency}

We first evaluate the efficiency of our algorithms. For each graph, we solve CFCM with cardinality constraint \(k=\abs{S}=20\). \tabref{tab:time} reports the running times of our algorithms and baselines. We present results for both of our algorithms with \(\epsilon\in\setof{0.3,0.2,0.15}\). Note that \algname{Exact} is infeasible for medium-scale graphs due to time-consuming matrix inversion, while \algname{ApproxGreedy} is infeasible for large-scale graphs, as its running time exceeds 27 hour.

\tabref{tab:time} shows that for every real-world graph, both of our algorithms outperform all the baselines in efficiency. Notably, the speed-up of our algorithms over \algname{ApproxGreedy} is more pronounced on denser graphs, such as \textit{buzznet} and \textit{Actor}. This observation aligns with our earlier theoretical analysis that the time complexity of \algname{ApproxGreedy} is dominated by number of edges rather than number of nodes.

Meanwhile, although both of our algorithms share the same theoretical upper bound of time complexity, the running time of \algname{SchurCFCM} is always lower than that of \algname{ForestCFCM}. This advantage lies in the introduction of Schur complement, letting \algname{SchurCFCM} sample spanning forests with more root nodes. During the sampling process, the random walker terminates once hitting the former trajectories. As \algname{RandomForest} is the most time-consuming part of \algname{ForestCFCM} and \algname{SchurCFCM}, the walker of \algname{SchurCFCM} is easier to terminate, significantly reducing running time.

\begin{figure}[htbp]
    \centering
    \includegraphics[width=\linewidth]{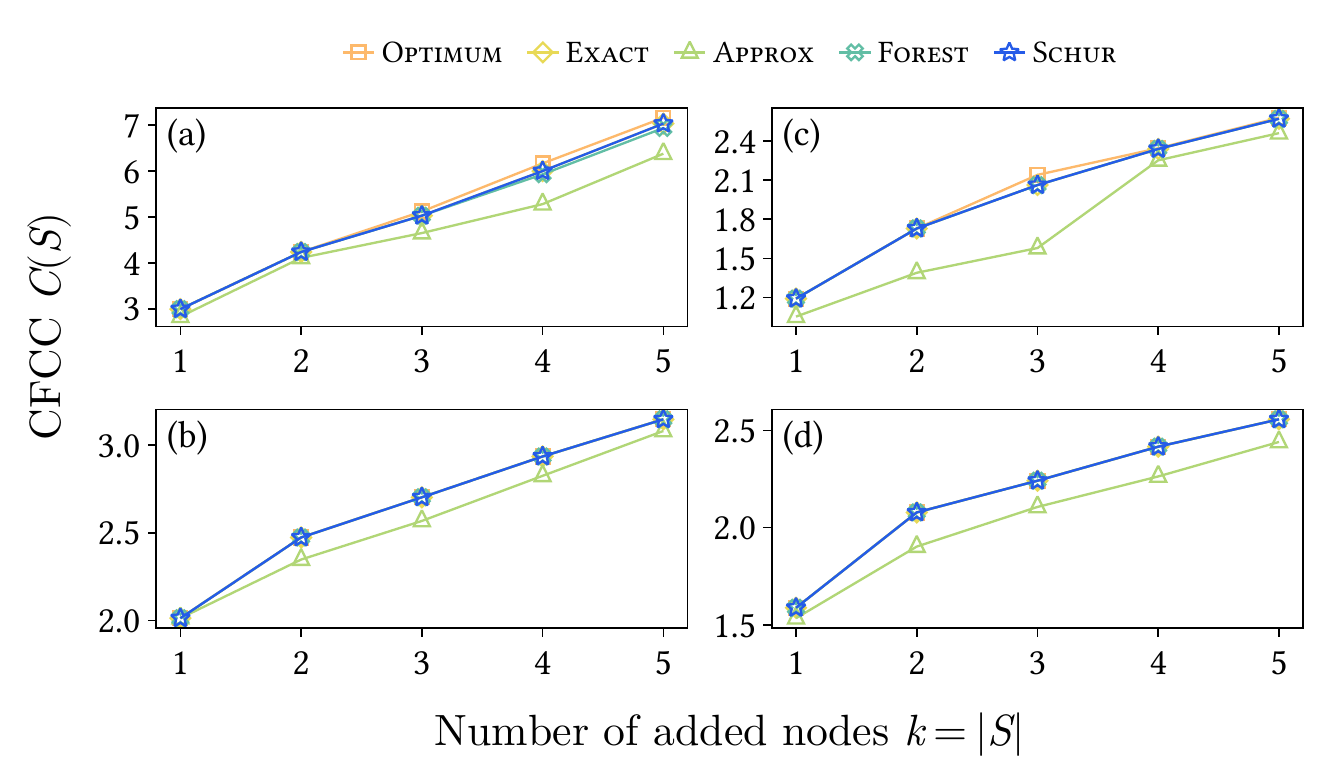}
    \caption{CFCC \(\cfcc{S}\) of node set \(S\) computed by different algorithms on four tiny-scale graphs: Zebra (a), Karate (b), Cont. USA (c) and Dolphins (d).}
    % \Description{CFCC computed by different algorithms on four tiny-scale graphs: Zebra (a), Karate (b), Cont. USA (c) and Dolphins (d).}
    \label{fig:optimum_effect}
\end{figure}

\subsubsection{Effectiveness}

We next evaluate the effectiveness of our algorithms. For tiny-scale graphs, the optimum solution of CFCM can be obtained by exhaustively examining all \(\binom{n}{k}\) selections. We compare the solutions from our algorithms with \(\epsilon=0.2\) and other baselines with the optimum solution on four tiny graphs~\cite{RoAh15}: \textit{Zebra} (23 nodes), \textit{Karate} (34 nodes), \textit{Cont. USA} (49 nodes) and \textit{Dolphins} (62 nodes). \figref{fig:optimum_effect} reports the results of different algorithms.

As shown by \figref{fig:optimum_effect}, the solutions provided by \algname{Exact} and our proposed algorithms are nearly identical, all very close to the optimum solutions. This indicates that the practical approximation ratios of our algorithms significantly outperform their theoretical guarantees. Furthermore, we observe that the approximation ratio of the state-of-the-art method \algname{ApproxGreedy} is lower than other methods. A similar phenomenon was reported in~\cite{LiPeShYiZh19}, which may be attributed to the inaccuracy of Laplacian solver for small graphs.

\begin{figure}[htbp]
    \centering
    \includegraphics[width=\linewidth]{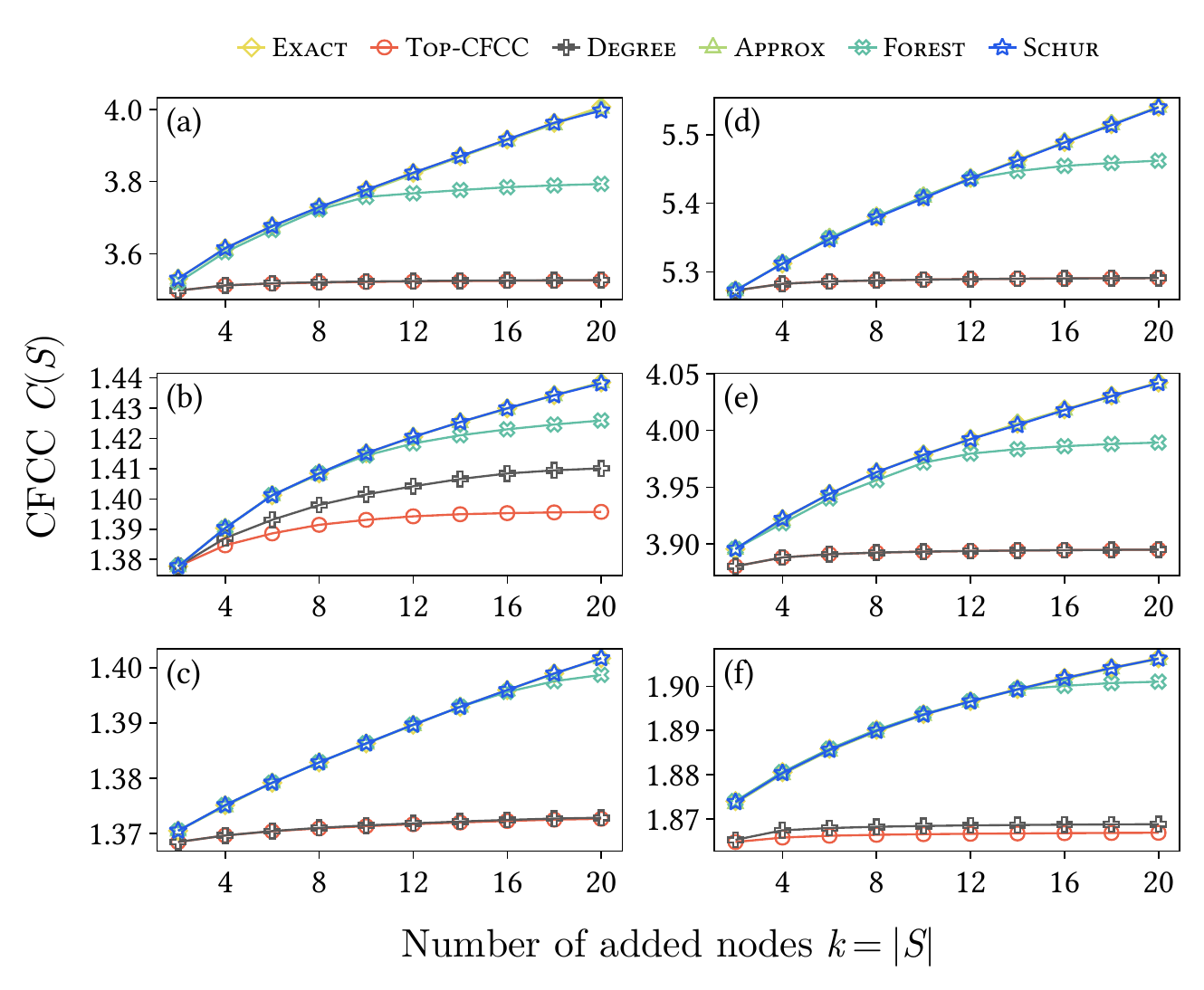}
    \caption{CFCC \(\cfcc{S}\) of node set \(S\) computed by different algorithms on small-scale graphs: Hamsterster (a), web-EPA (b), Routeviews (c), soc-PagesGov (d), Astro-Ph (e) and EmailEnron (f).}
    \label{fig:exact_effect}
\end{figure}

\begin{figure}[htbp]
    \centering
    \includegraphics[width=\linewidth]{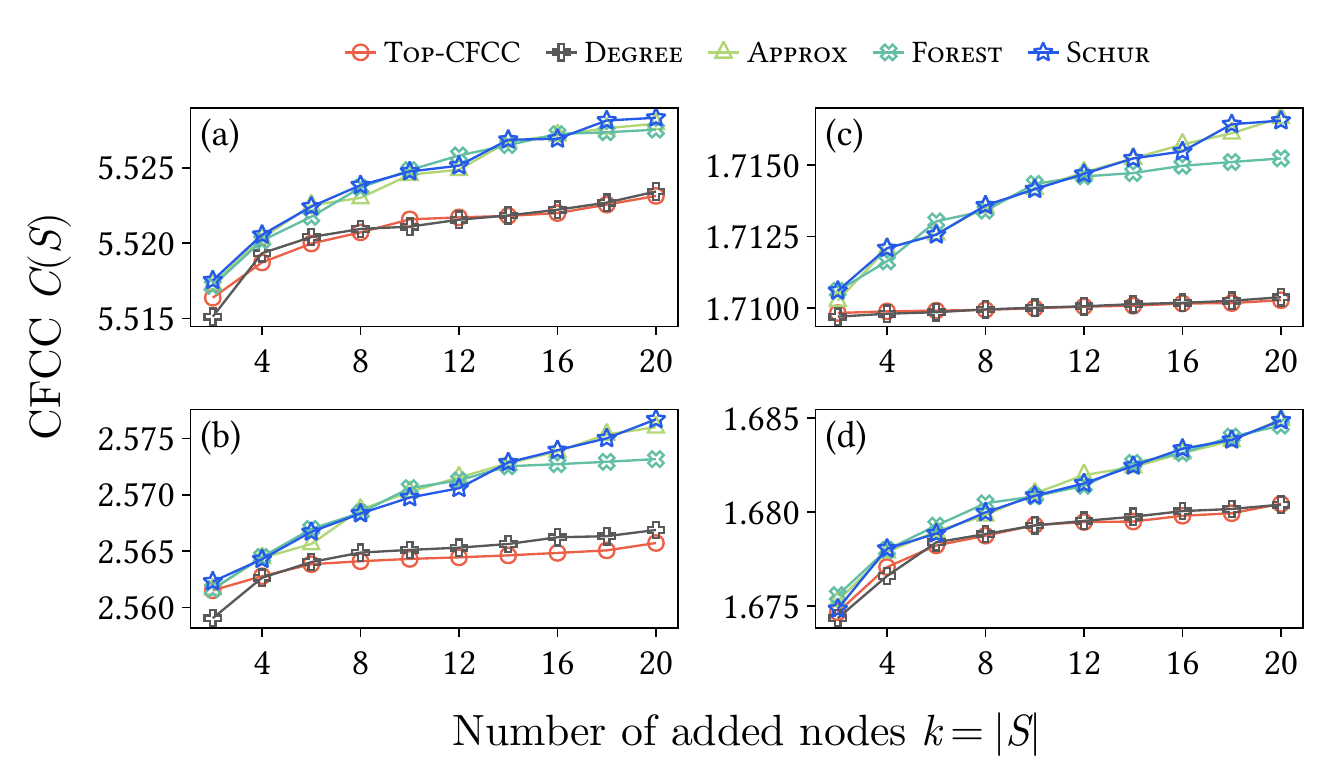}
    \caption{CFCC \(\cfcc{S}\) of node set \(S\) by different algorithms on large-scale graphs: Livemocha (a), WordNet (b), Gowalla (c) and com-DBLP (d).}
    \label{fig:approx_effect}
\end{figure}

We further evaluate the performance of our algorithms against heuristic approaches \algname{Degree} and \algname{Top-CFCC} described in \secref{subsec:exp-setting}. These methods are executed along with four greedy algorithms on six small-scale graphs. For large-scale networks where direct computation of CFCC and CFCM becomes infeasible, we employ the conjugate gradient method~\cite{Sa03} to examine approximate solutions, excluding \algname{Exact}. The results presented in \figref{fig:exact_effect} and \figref{fig:approx_effect} reveal three key observations. First, while \algname{ForestCFCM} achieves superior effectiveness in initial iterations, its CFCC maximization rate is eventually surpassed by other methods. Second, \algname{SchurCFCM} consistently delivers the most effective solutions throughout all iterations, benefiting from the Schur complement introduced in \lemref{lem:schur-sublap}. Third, the heuristic method \algname{Top-CFCC} shows comparable or inferior performance to \algname{Degree}, suggesting that single-node centrality rankings alone cannot effectively identify crucial node groups.

\subsection{Influence of Varying Error Parameter}

Our analysis of algorithmic efficiency and solution quality reveals significant sensitivity to the error parameter \(\epsilon\). We systematically investigate this relationship by varying \(\epsilon\) within \([0.15, 0.4]\) and measuring algorithm performance across multiple real-world networks. The evaluation focuses on two metrics: computational time across different graph sizes and relative differences of maximized CFCC compared to \algname{Exact}.

\begin{figure}[htbp]
    \centering
    \includegraphics[width=\linewidth]{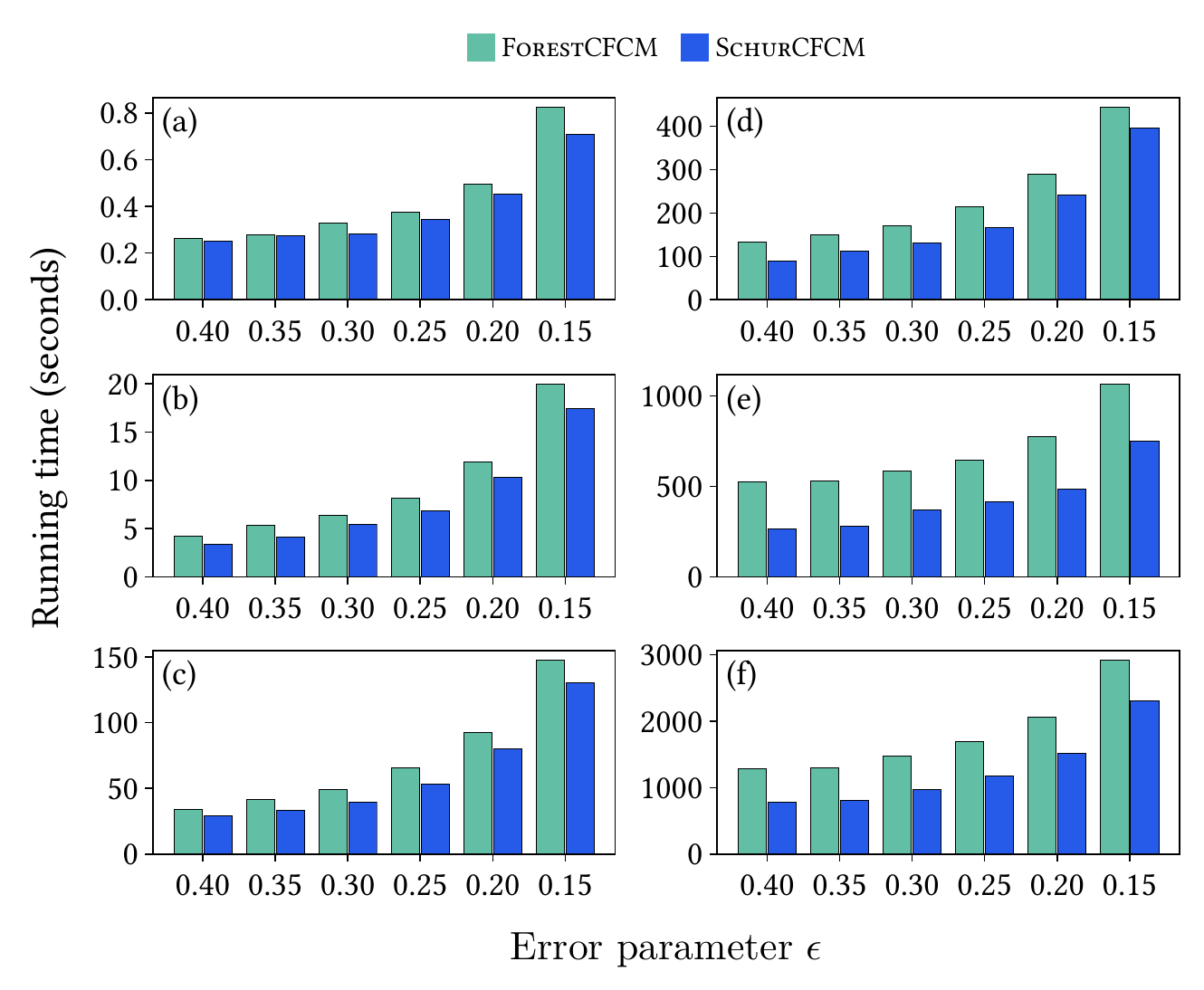}
    \caption{Running time of different algorithms with varying error parameter \(\epsilon\) on real-world graphs: Euroroads (a), soc-PagesGov (b), EmailEnron (c), com-DBLP (d), Skitter (e) and sc-rel9 (f).}
    \label{fig:realistic_time}
\end{figure}

\subsubsection{Effect on efficiency}

We first assess the impact of varying error parameter on the efficiency of our algorithms. \figref{fig:realistic_time} demonstrates the computational time scaling of our algorithms with various \(\epsilon\). Both of our algorithms exhibit similar growth patterns that align with the factor \(\epsilon^{-2}\) of their complexity. Notably, \algname{SchurCFCM} shows increasing efficiency advantages at smaller \(\epsilon\) values, particularly when \(\epsilon \leq 0.2\). This enhanced performance further validates the improvements of introducing Schur complement by \algname{SchurCFCM}, as smaller \(\epsilon\) leads to more sampled spanning forests.

\begin{figure}[htbp]
    \centering
    \includegraphics[width=\linewidth]{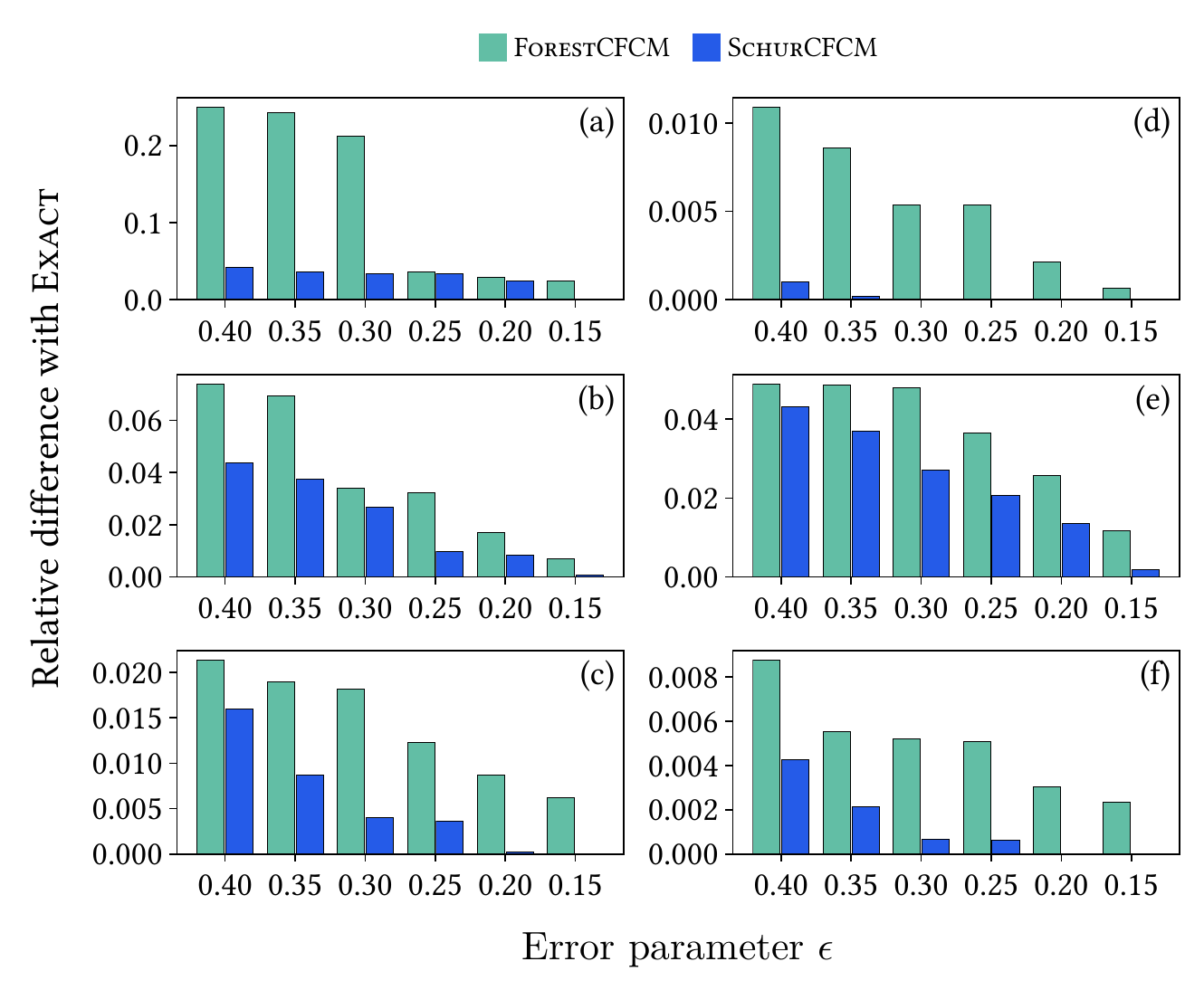}
    \caption{Relative difference of different algorithms with varying error parameter \(\epsilon\) on small-scale graphs: Facebook (a), GR-QC (b), web-EPA (c), Routeviews (d), HEP-Th (e) and CAIDA (f).}
    \label{fig:realistic_cfcc}
\end{figure}

\subsubsection{Effect on effectiveness}

We next analyze the impact of varying \(\epsilon\) on the effectiveness of our algorithms. The results are presented in \figref{fig:realistic_cfcc}. As displayed in \figref{fig:realistic_cfcc}, \algname{SchurCFCM} maintains superior CFCC maximization across all tested \(\epsilon\) values. Although the relative differences of our algorithms with large \(\epsilon\) are not ideal, reducing \(\epsilon\) to \(0.2\) or \(0.15\) significantly decreases their differences to negligible levels. The observed sensitivity to \(\epsilon\) variations demonstrates effective parameter governance in both algorithms, with improvements of solution quality saturating beyond \(\epsilon = 0.2\).

\section{Related Work}

\textbf{Computation of resistance distance.}
As a fundamental metric, resistance distance has found wide applications in the data management community, such as recommendation systems~\cite{YiCuLiYaCh12} and graph embedding systems~\cite{QiDhTaPeWa21}. Numerous algorithms have been proposed to compute resistance distance efficiently~\cite{PeLoYoGo21,YaTa23,ChGaPeSaSaWa23,LiZhLiDaChWa24}. However, these approaches are based on the interpretations of resistance distance expressed in~\eqref{eq:resist-sublap-diag}, rather than the marginal gain represented in~\eqref{eq:def-gain-subseq}. Therefore, existing methods for computing resistance distance cannot be directly applied to solving CFCM.

\textbf{Other node selection problems.}
There exist various importance measures of a group of nodes, based on graph structure~\cite{MaTsUp16} or dynamic processes~\cite{LiYuHuCh14}. As the criteria for identifying crucial node groups are application-dependent~\cite{GhTeLeYa14}, many previous studies have focused on selecting \(k\) nodes to optimize related quantities. Recently, influence maximization (IM) has garnered wide interest, with information spread as the optimized quantity. However, the predominant processes for information diffusion are independent cascade~\cite{ChRaMaDe23,FePaCaVa23,GuFeZhWa23,HuLiBaSu22} and linear threshold~\cite{HuZhLi23,ZhHuSuLiXiTa23}, both of which differ greatly from the electrical network model. Consequently, existing approaches for IM cannot be transferred to solving CFCM.

\textbf{Discussion of edge selection problems.}
Beyond node selection, significant efforts have been devoted to the problem of adding \(k\) edges connecting to a given node to maximize its centrality. This problem has been studied via diverse centrality measures, including betweenness centrality~\cite{HoMoSo18}, information centrality~\cite{ShYiZh18} and absorbing random-walk centrality~\cite{AdWaGi23}. In contrast to single-node scenarios, the problem of maximizing the centrality of a node group by adding \(k\) edges connecting nodes within the group has been studied for only a few centralities, such as group betweenness~\cite{MeSiSiBaSw18}. Previous works have not solved the edge selection problem for maximizing CFCC, which presents an opportunity for future research.

\section{Conclusion}

For real-world graphs with \(n\) nodes, we developed two greedy algorithms to approximately maximize the current flow closeness centrality of node groups with cardinality constraint \(k\ll n\). Our first algorithm, \algname{ForestCFCM}, is based on spanning forest sampling and leverages an adaptive sampling technique. \algname{ForestCFCM} exhibits a nearly-linear time complexity with respect to \(n\), outperforming the state-of-the-art method. To further improve efficiency and effectiveness, we proposed our second algorithm \algname{SchurCFCM}, which additionally utilizes the Schur complement. Both of our greedy algorithms achieve a guaranteed approximation factor of \(1-\frac{k}{k-1}\frac{1}{\rme}-\epsilon\) for any error parameter \(0<\epsilon<1\). Numerical results demonstrate that both algorithms are significantly faster than the state-of-the-art method, scaling to real-world graphs with 3 million nodes. Notably, \algname{SchurCFCM} outperforms all other approaches in terms of both efficiency and effectiveness.

\section*{Acknowledgements}

This work was supported by the National Natural Science Foundation of China (Nos. 62372112 and 61872093).

\bibliographystyle{IEEEtran}
\balance
\bibliography{
    ref/applic.bib,
    ref/resist.bib,
    ref/related_work.bib,
    ref/misc.bib
}

\end{document}